\documentclass{pasj01}
\draft

\Received{}
\Accepted{}

\newcommand\figr[1]{\ref{fig:#1}}
\newcommand\tabr[1]{\ref{tab:#1}}
\newcommand\figref[1]{figure~\ref{fig:#1}}
\newcommand\tabref[1]{table~\ref{tab:#1}}
\newcommand\Figref[1]{Figure~\ref{fig:#1}}

\usepackage{afterpage}
\usepackage{lineno} 

\graphicspath{{fig/}}

\begin{document}

\title{ALMA Band 6 high-resolution observations of the transitional disk around SY Cha}

\author{Ryuta \textsc{Orihara},\altaffilmark{1}
        Munetake \textsc{Momose},\altaffilmark{1,$^{*}$}
        Takayuki \textsc{Muto},\altaffilmark{2}
        Jun \textsc{Hashimoto},\altaffilmark{3}
        Hauyu Baobab \textsc{Liu},\altaffilmark{4}
        Takashi \textsc{Tsukagoshi},\altaffilmark{5}
        Tomoyuki \textsc{Kudo},\altaffilmark{6}
        Sanemichi \textsc{Takahashi},\altaffilmark{7}
        Yi \textsc{Yang},\altaffilmark{3}
        Yasuhiro \textsc{Hasegawa},\altaffilmark{8}
        Ruobing \textsc{Dong},\altaffilmark{9}
        Mihoko \textsc{Konishi}\altaffilmark{10}
        and Eiji \textsc{Akiyama}\altaffilmark{11}}

\altaffiltext{1}{College of Science, Ibaraki University, 2-1-1 Bunkyo, Mito, Ibaraki 310-8512, Japan}
\altaffiltext{2}{Division of Liberal Arts, Kogakuin University, 1-24-2, Nishi-Shinjuku, Shinjuku-ku, Tokyo, 163-8677, Japan}
\altaffiltext{3}{Astrobiology Center, National Institutes of Natural Sciences, 2-21-1 Osawa, Mitaka, Tokyo 181-8588, Japan}
\altaffiltext{4}{Institute of Astronomy and Astrophysics, Academia Sinica, 11F of Astronomy-Mathematics Building, AS/NTU No. 1, Sec. 4, Roosevelt Road, Taipei 10617, Taiwan, ROC}
\altaffiltext{5}{Faculty of Engineering, Ashikaga University, Ohmae-cho 268-1, Ashikaga, Tochigi, 326-8558, Japan}
\altaffiltext{6}{SubaruTelescope, National Astronomical Observatory of Japan, 650 North A'ohoku Place, Hilo, HI 96720 USA}
\altaffiltext{7}{National Astronomical Observatory of Japan, 2-21-1 Osawa, Mitaka, Tokyo 181-8588, Japan}
\altaffiltext{8}{Jet Propulsion Laboratory, California Institute of Technology, Pasadena, CA 91109, USA}
\altaffiltext{9}{Department of Physics \& Astronomy, University of Victoria, Victoria, BC, V8P 5C2, Canada}
\altaffiltext{10}{Faculty of Science and Technology, Oita University, 700 Dannoharu, Oita 870-1192, Japan}
\altaffiltext{11}{Department of Engineering, Niigata Institute of Technology, 1719 Fujihashi, Kashiwazaki, Niigata 945-1195, Japan}

\email{munetake.momose.dr@vc.ibaraki.ac.jp}
\KeyWords{protoplanetary disks $-$ stars: individual (SY Cha) $-$ stars: pre-main sequence $-$ submillimeter: planetary systems}

\maketitle

\begin{abstract}
In this study, we reported the results of high-resolution (0.14 arcsec) Atacama Large Millimeter/submillimeter Array (ALMA) observations of the 225 GHz dust continuum and CO molecular emission lines from the transitional disk around SY Cha.
Our high-resolution observations clearly revealed the inner cavity and the central point source for the first time.
The radial profile of the ring can be approximated by a bright narrow ring superimposed on a fainter wide ring.
Furthermore, we found that there is a weak azimuthal asymmetry in dust continuum emission.
For gas emissions, we detected $\rm{}^{12}CO$(2$-$1), $\rm{}^{13}CO$(2$-$1) and $\rm{}C^{18}O$(2$-$1), from which we estimated the total gas mass of the disk to be $2.2\times10^{-4}M_\odot$, assuming a CO/H$_2$ ratio of $10^{-4}$.
The observations showed that the gas is present inside the dust cavity.
The analysis of the velocity structure of the $\rm{}^{12}CO$(2$-$1) emission line revealed that the velocity is distorted at the location of the dust inner disk, which may be owing to warping of the disk or radial gas flow within the cavity of the dust disk. 
High-resolution observations of SY Cha showed that this system is composed of a ring and a distorted inner disk, which may be common, as indicated by the survey of transitional disk systems at a resolution of $\sim$0.1~arcsec.
\end{abstract}

\section{Introduction}
Protoplanetary disks around young stars are composed of gas and dust, and they are believed to be the birth place of planets.
After the formation of a planet, the gravitational interaction between the disk and planet is expected to excite various structures in the disk.
Recently, many examples of complex structures in protoplanetary disks have been identified by Atacama large millimeter/submillimeter array (ALMA) observations (e.g., \cite{fukagawa2013}; \cite{ninke2013}; \cite{alma2015}; \cite{andrews2016}). 
Among them, disks with a depletion of dust close to the central star have been identified and are called transitional disks (\cite{ninke2018}).
It is believed that the transitional disks are evolved protoplanetary disks;
however, there are also some signs that these disks pass through a separate evolutionary process (\cite{pinilla2018}).
Transitional disks were originally identified as young stellar objects (YSOs) with low near-IR excesses (1$-$5~$\rm{\mu m}$), but with significant excesses at mid-IR (5$-$20~$\rm{}\mu m$) and far-IR wavelengths in their spectral energy distributions (SEDs) (\cite{Calvet2002}; \cite{Calvet2005}).

\begin{table}[t]
\tbl{Properties of SY Cha}{%
\begin{tabular}{l|l}    
    \hline
    Parameter&Value\\
    \hline
    Stellar Mass $M_{\star}$&$0.78\ M_{\odot}$\footnotemark[ (1)]\\
    Mass accretion rate $\dot{M_\star}$&$3.89\times10^{-10}\ M_{\odot}/\rm{yr}$\footnotemark[ (1)]\\
    Stellar Age & 3 Myr\footnotemark[ (2,3)]\\
    Stellar Radius $R_{\star}$&$1.58\ R_{\odot}$\footnotemark[ (4)]\\
    Stellar Limunosity $L_{\star}$&$0.465\ L_\odot$\footnotemark[ (4)]\\
    Effective Temperature $T_{\rm{}eff}$&$3792\ \rm{K}$\footnotemark[ (4)]\\
    Distance $d$&$180.7\ \rm{pc}$\footnotemark[ (5)]\\
    Spectral type &K5Ve\footnotemark[ (6)]\\
    Proper Motion $\Delta{\rm{Ra.}}$&$-23.738\ \rm{mas/yr}$\footnotemark[ (5)]\\
    Proper Motion $\Delta{\rm{Dec.}}$&$2.804\ \rm{mas/yr}$\footnotemark[ (5)]\\
    \hline
\end{tabular}}\label{tab:sycha}
\begin{tabnote}
\footnotemark[(1)]\citet{manara2017}, 
\footnotemark[(2)]\citet{siess1997},
\footnotemark[(3)]\citet{siess2000},
\footnotemark[(4)]\citet{gaia2018},
\footnotemark[(5)]\citet{gaiaedr3},
\footnotemark[(6)]\citet{gaia2015}
\end{tabnote}
\end{table}

\begin{figure}[t]
    \begin{center}
    \includegraphics[width=.45\textwidth]{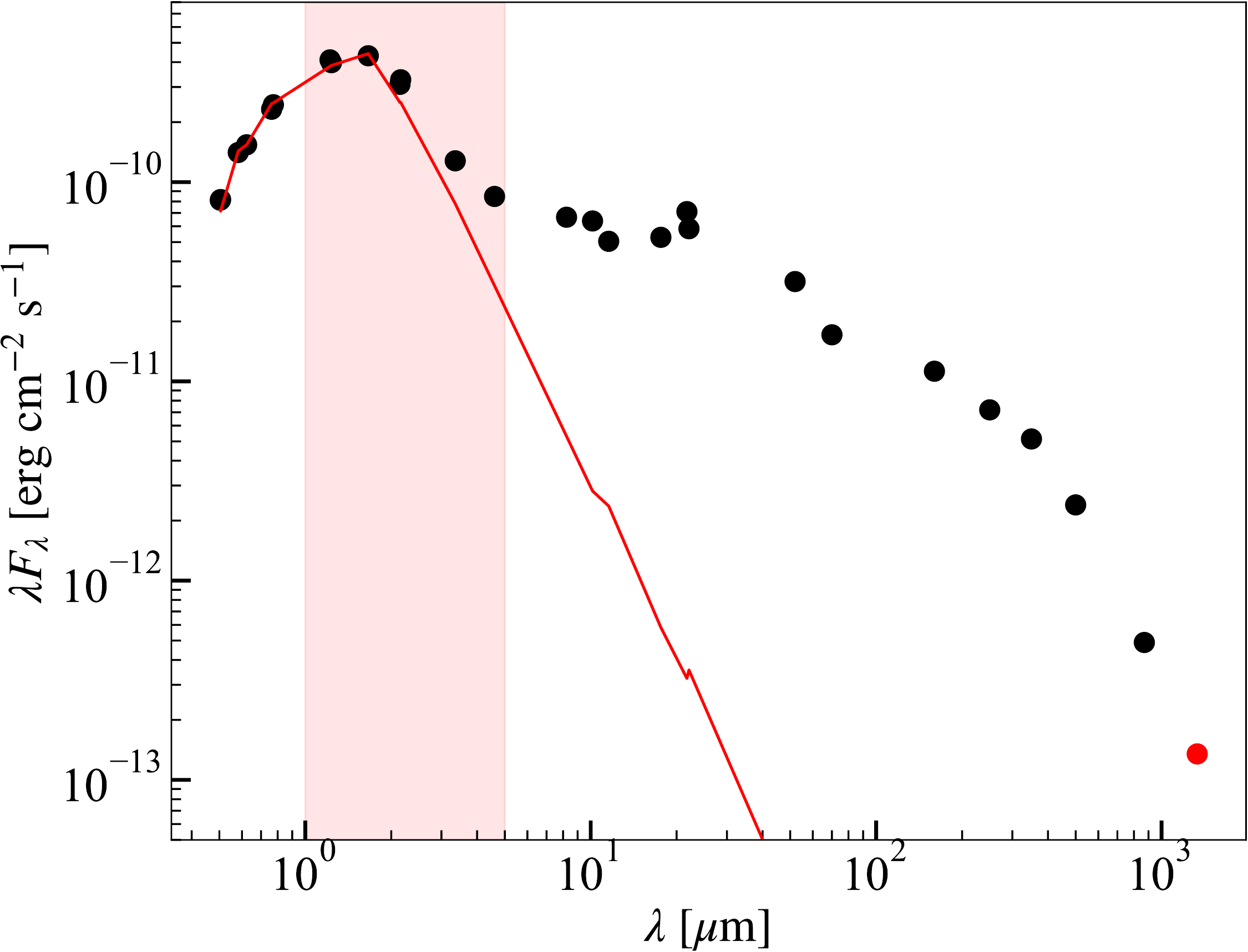}%
    \end{center}
    \caption{Spectral energy distribution (SED) of SY Cha (\cite{DENIS2000}; \cite{2MASS2003}; \cite{WISE2010}; \cite{AKARI2010}; \cite{IRAS1984}; \cite{Herschel2012}; and \cite{APEX2011}).
    The red line indicates the photospheric emission estimated by VOSA (\cite{vosa2008}). 
    The red circle is the flux information obtained by our observation.
    The red filled area indicate near-IR wavelengths (1$-$5~$\rm{}\mu m$).}
    \label{fig:SED}
\end{figure}

The cavity seen in transitional disks can be formed by an embedded planet or multiple planets (\cite{dong15gap}), which can create local maxima in the gas pressure where large dust grains are trapped (\cite{pinilla2012}). 
A system harboring an inner dust disk with a size of a few au is particularly intriguing because such a disk may be deeply related to the formation of terrestrial planets in the vicinity of the central star (\cite{francis2020}).
A misalignment between the inner and outer disks has also been confirmed from a distortion of the velocity field in the disk and can be caused by a gravitational interaction with the planet (e.g., \cite{mayama2018}; \cite{casassus2015}).
\citet{francis2020} analyzed ALMA archival data of the continuum emissions from 38 transitional disks and found that at least 18 transitional disks in their sample harbored an inner disk in the central cavity. 
Therefore, detailed studies on structures and evolution of transitional disks are important for better understanding how planet formation progresses in the disk. 
Also, the transitional disks will be promising targets for direct imaging of planets in near future.

Recent ALMA high-resolution observations have revealed an inner ring with a radius of $\sim 4$~au in the disk around DM Tau, which has been classified as a transitional disk object (\cite{espaillat2011}; \cite{Kudo2018}).
This result shows that whether there is an additional component in the central gap of a transitional disk may not be judged solely from its SED. 
Moreover, the near-infrared excess in a transitional disk may not be correlated with the mass of mm-size dust in the inner disk (\cite{francis2020}).
Thus, high-resolution observations that reveal the actual distributions of gas and dust at a small scale are necessary to understand more about the origin of transitional disks.

In this paper, we report the results of high-resolution ALMA observations of the disk around SY Cha.
SY Cha is a T Tauri star with the mass of 0.78 $M_\odot$ (\cite{manara2017}).
The stellar age is estimated to be $\sim$3~Myr (\cite{siess1997}; \cite{siess2000}), and it exhibits a mass accretion rate of $\dot{M}_{\star} = 3.89 \times 10^{-10}$~$M_\odot$/yr (\cite{manara2017}).
Other properties are summarized in \tabref{sycha}.
The SED of SY Cha (\figref{SED}) does not show a significant excess in the near-infrared wavelength regime.
However, for the first time, our highly sensitive and high spatial resolution observations of the circumstellar environment revealed the presence of an inner disk around SY Cha.
This paper provides an example of observation of a transitional disk with an inner disk.

The remainder of this paper is structured as follows. We present the observational setup and data reduction in section~2, and then we report the analysis results in section~3.
In section~4, we discuss the causes of the velocity distortion detected by our observations.
Finally, we present conclusions in section~5.


\begin{table*}[t]
\tbl{Observational Setup}{%
\begin{tabular}{l|lll} 
    \hline
                           & EC data & CC data 1 & CC data 2 \\
    \hline
    Obs. Date              & 22 June 2019          & 19 August 2019           & 25 September 2019           \\
    Configuration          & C43-9   & C43-7    & C43-6 \\
    Number of Antennas         & 48               & 45                   & 50                   \\
    Max. Baseline [km]      & 16.2               & 3.6                 & 2.6                 \\
    Min. Baseline [m]      & 83.1                & 41.4                 & 15.1                 \\
    On-source Time [min]   & 72.75               & 15.25                & 15.25                \\ 
    \hline
    \shortstack{Calibrators\\{}\\{}\\{}\\{}\\{}\\{}\\{}\\{}}
                     & \shortstack[l]{\\J1107-4449\footnotemark[ 1,2,3,4]\\{}\\
                                   J1617-5848\footnotemark[ 1,2,3,4]\\{}\\
                                   J1145-6954\footnotemark[ 4]\\{}\\
                                   J1148-7819\footnotemark[ 5]\\{}\\
                                   J1155-8101\footnotemark[ 6]}
                     & \shortstack[l]{\\J1107-4449\footnotemark[ 1,2,3,4]\\{}\\
                                   J1058-8003\footnotemark[ 4,5]\\{}\\
                                   J1155-8101\footnotemark[ 6]\\{}\\{}\\{}\\{}}
                     & \shortstack[l]{\\J0635-7516\footnotemark[ 1,2,3,4]\\{}\\
                                   J1058-8003\footnotemark[ 5]\\{}\\
                                   J1155-8101\footnotemark[ 6]\\{}\\{}\\{}\\{}}                                \\  \hline
    CASA pipeline version           & 42254M              & 42866M               & 42866M      \\ 
    \hline
\end{tabular}}\label{tab:obsset}
\begin{tabnote}
\footnotemark[1]Absolute flux calibrator, \footnotemark[2]Atmosphere calibrator, \footnotemark[3] Bandpass calibrator, \footnotemark[4] Pointing calibrator,\\ 
\footnotemark[5] Complex gain calibrator, \footnotemark[6] Check source
\end{tabnote}
\end{table*}

\begin{table*}[t]
\tbl{Correlator Setup}{%
\begin{tabular}{l|cccc} 
    \hline
    & Continuum\footnotemark[ 1] & $\rm{}^{12}CO$(2$-$1) & $\rm{}^{13}CO$(2$-$1) & $\rm{}C^{18}O$(2$-$1)\\ 
    \hline
    Central Frequency [GHz] & 216.984,\ 232.983 & 230.521 & 220.383 & 219.544 \\
    Number of Channel       & 128,\ 128   & 3840 & 1920 & 1920   \\
    Band Width [GHz]        & 2,\ 2   & 0.94 & 0.47 & 0.47 \\
    Velocity width [km/s]   & 21.6,\ 20.1\, &0.317&0.332&0.333\\
    Rest Frequency [GHz]  & - &230.538&220.399&219.560\\
    \hline
\end{tabular}}\label{tab:correlator}
\begin{tabnote}
\footnotemark[1](Lower sideband, Upper sideband)
\end{tabnote}
\end{table*}

\begin{table*}[t]
\tbl{Imaging Parameters}{%
\begin{tabular}{l|ccccc} 
    \hline
                            & \shortstack[c]{{}\\Continuum\\{}} & \shortstack[c]{{}\\Continuum\\ taper} & \shortstack[c]{{}\\$\rm{}^{12}CO$\\{}} & \shortstack[c]{{}\\$\rm{}^{13}CO$\\{}} & \shortstack[c]{{}\\$\rm{}C^{18}O$\\{}} \\
    \hline
    Channel Width [km/s]    & 5330 & 5330 & 0.7  & 0.7 & 0.7 \\
    Number of Channels    & 1         & 1         & 25  & 25  & 25\\
    Density Weighting (Robust)
    & 0.5 & 0.5 & 0.5 & 0.5 & 0.5\\
    Gaussian taper       & - & 1.5 M$\lambda$ & 1.5 M$\lambda$ & 0.3 M$\lambda$ & 0.3 M$\lambda$\\
    \shortstack[l]{Beam size\\Beam Position Angle\\{}}               
    &\shortstack{{}\\$0.040^{\prime\prime}\times0.022^{\prime\prime}$\\$17.5^\circ$\\{}}
    &\shortstack{{}\\$0.120^{\prime\prime}\times0.083^{\prime\prime}$\\$6.3^\circ$\\{}}
    &\shortstack{{}\\$0.128^{\prime\prime}\times0.085^{\prime\prime}$\\$4.7^\circ$\\{}}
    &\shortstack{{}\\$0.356^{\prime\prime}\times0.349^{\prime\prime}$\\$-37.2^\circ$\\{}}
    &\shortstack{{}\\$0.356^{\prime\prime}\times0.348^{\prime\prime}$\\$-29.5^\circ$\\{}}\\
    RMS Noise [mJy/beam]    & 11.2$\times10^{-3}$  & 14.9$\times10^{-3}$ & 1.34 & 2.18 & 1.74 \\ 
    Peak Channel  ($V_{\rm{}LSRK}$)[km/s]     & -  & - & 2.0 & 6.2 & 6.2\\
    Peak Intensity [mJy/beam]  & 0.347 & 0.962 & 20.9 & 52.8 & 18.4\\
    SNR & 31.0 & 64.6 & 15.6 & 24.2 & 10.6\\
    Total Flux Density [mJy] & 57.7 & 55.9 & - & - & - \\
    Integrated Flux Density [Jy $\cdot$ km/s] & - & - & 7.9 & 1.4 & 0.3\\
    \hline
\end{tabular}}\label{tab:imageparam}
\end{table*}

\section{Observations and data reduction} 

SY Cha was observed with ALMA in Band 6 during Cycle 6 (2018.1.00689S, PI: Muto).
The observations were conducted using an extended configuration (EC) on June 22, 2019, and by compact configuration (CC) on August 19 and September 25, 2019, with the spectral windows (SPW) covering $\rm{}^{12}CO$(2$-$1), $\rm{}^{13}CO$(2$-$1), $\rm{}C^{18}O$(2$-$1) and continuum emission, as summarized in tables \tabr{obsset} and \tabr{correlator}.

The visibility data were reduced and calibrated using the Common Astronomy Software Applications (CASA) package (\cite{macmullin2007}). 
We used the CASA pipeline (EC data: CASA 5.40-70 42254M, CC data: CASA 5.6.1-8 42866M) to calibrate the data, as well as to flag bad data.

Two spectral windows for continuum observations with a bandwidth of 2 GHz, as well as three spectral windows for line observations, which were separated into continuum and line by the \texttt{uvcontsub} task in the CASA tools (\tabref{correlator}), were combined and averaged to obtain continuum data centered at 225 GHz.
The continuum image was reconstructed by multi-scale CLEAN (\cite{cornwell2008}) of CASA (version 6.4.4.31) with natural weighting and \texttt{scales parameters} of [0, `beam size', `$2 \times$beam size', `$3 \times$beam size', `$4 \times$beam size', `$5 \times$beam size'].

We performed self-calibration in phase for CC visibility using a model of the disk around SY Cha created by the \texttt{tclean} task in the CASA tools.
Since the EC observation did not detect the emission with sufficient sensitivity, no self-calibration was performed for it.
It is known that only correcting for proper motion is sometimes inadequate and the alignment of ALMA data taken with different execution blocks is required (\cite{hashimoto2021}).
To correct for misalignment of the phase center between observations, the phase centers of all observations were set at the position where the root mean square (RMS) of the imaginary part of the visibility was minimized by the \texttt{phaseshift} and \texttt{fixplanets} tasks in the CASA tools, taking advantage of the fact that the disk is nearly axisymmetric (\cite{kanagawa2021}).
The real and imaginary parts of the visibility correspond to the symmetric and asymmetric components, respectively, of the image with respect to 180$^\circ$ rotation.
Here, because it is not necessarily true that the asymmetry would behave similarly on different spatial scales, the above process was limited to a spatial frequency range of $300\sim 1000\rm{}k\lambda$, which was a common sample for each observation.
The final round of self-calibration in phase was then performed with \texttt{combine} ``spw, scan'' and \texttt{solint} ``inf'' for concatenated visibility. 
This corrects for the slight difference of the phase center between observations. 
Finally, we obtained a dust continuum image with a beam size of $0.040^{\prime\prime}\times0.022^{\prime\prime}$ and an RMS noise of $11.2\ \rm{\mu Jy/beam}$.
However, this image is not sensitive to weakly extended emissions, so we also prepared the tapered image with a beam size of $0.120^{\prime\prime}\times0.083^{\prime\prime}$ and an RMS noise of $14.9\ \rm{\mu Jy/beam}$.

Continuum components were subtracted from all line data in the visibility domain with the task \texttt{uvcontsub} in the CASA tools.
Before imaging the line emission, the solution tables obtained from the self-calibration of the continuum data were applied to the line data.
Channel maps with a velocity width of 0.7 km/s were created by Hogbom CLEAN in a velocity range of $V_{\rm{LSRK}}= -4.3–12.5$~km/s. Subsequently, we set the density weight and scale parameters similar to the continuum imaging.  Because $\rm{}^{12}CO$(2$-$1) emission was detected in the EC data, we performed imaging using the concatenated visibility.
The beam size and RMS noise of each channel map are presented in \tabref{imageparam}. 
To enhance the SNR, we applied a Gaussian uvtaper of $1.5\ \rm{}M\lambda$ on the $\rm{}^{12}CO$, and $0.3\ \rm{}M\lambda$ on the $\rm{}^{13}CO$ and $\rm{}C^{18}O$.
The beam size of the highest-resolution image is $0.128^{\prime\prime}\times0.085^{\prime\prime}$.

We created the moment maps from each channel map using the \texttt{bettermoments} package (\cite{better2018}).
Here, the channel maps were applied to the Keplerian mask generated with the \texttt{keplerian\_mask} package (\cite{teague2020}) before creating the moment map.
This package can calculate masks based on the expected emitting surface of a Keplerian disk.
The calculation considered the stellar mass $M_\star=0.78\ M_\odot$ (\cite{manara2017}), distance to the star $d = 180.7$ pc (\cite{gaiaedr3}), and applied the position angle (PA) and inclination obtained from our continuum observations (see section 3.1.1). 
In addition, we assumed the height of the emitting surface relative to the radius to be 0.13 from the estimation of scale height (see section 3.3.1).
The mask was convolved with a 2D Gaussian that has the same size as the synthesized beam of each channel map.
When creating moment 1 and 2 maps, the clipping levels were set to 3$\sigma$ (1$\sigma$ is the RMS of each channel map).

\section{Data analysis and results}

\subsection{Synthesized images}

\subsubsection{225GHz dust continuum}

\begin{figure*}[t!]
\begin{center}
\includegraphics[width=\textwidth]{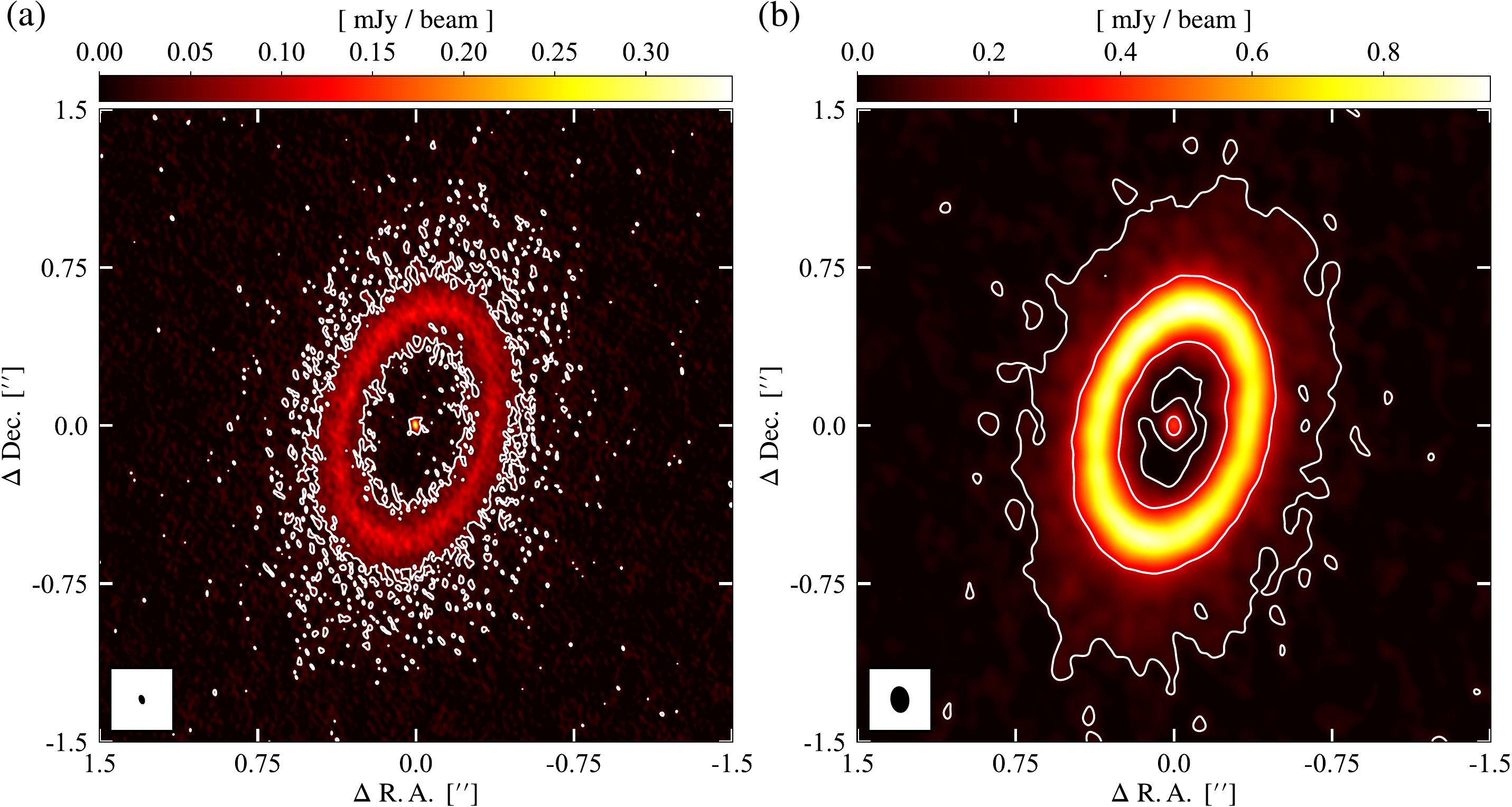}%
\end{center}
\caption{225 GHz dust continuum images around SY Cha. 
    (a) The synthesized beam size ($0.040^{\prime\prime}\times0.022^{\prime\prime}$) is shown at the bottom left.
    The white contour denotes the $3\sigma$, where the $1\sigma$ level is the RMS noise (= $11.2\ \rm{\mu Jy\ beam^{-1}}$).
    (b) The image was made by adding uvtaper to the \texttt{tclean} parameter.
    The synthesized beam size ($0.120^{\prime\prime}\times0.083^{\prime\prime}$) is shown at the bottom left.
    The white contour denotes the $3\sigma,\ 20\sigma$, where the $1\sigma$ level is the RMS noise (= $14.9\ \rm{\mu Jy\ beam^{-1}}$).
    }
    \label{fig:dust_continuum}
    \end{figure*}

\begin{figure*}[t!]
\begin{center}
\includegraphics[width=\textwidth]{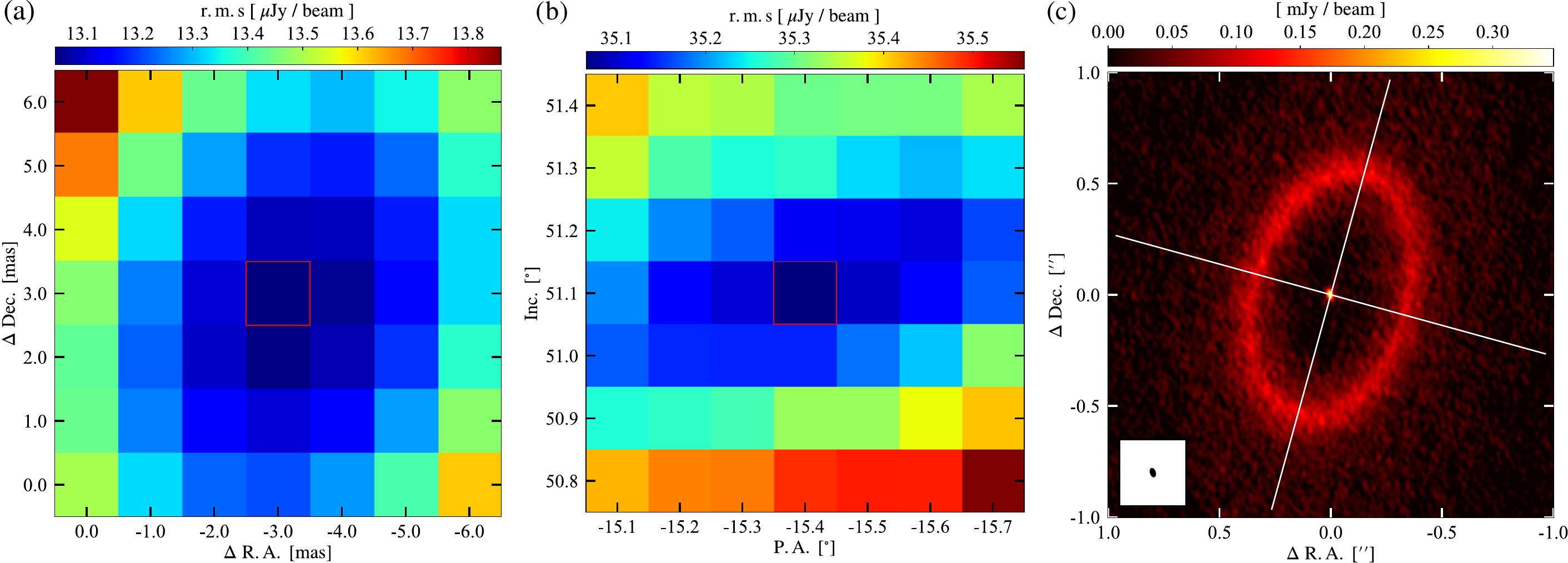}%
\end{center}
\caption{
    (a) RMS of the images synthesized with only the imaginary part of the visibility by shifting in steps of 1 mas in R.A. and Dec directions.
    The axes show a shift from the phase center of the point source.
    The RMS is calculated in the region with a radius of 2 arcsec.
    The phase center shifted by $-3$ mas in the RA direction and 3~mas in the Dec direction minimizes the RMS of the image made from the visibility with real part $=$ 0 (red box).
    (b) RMS of difference between the image and its $90^\circ$-rotation.
    The images were synthesized with the deprojected visibility by changing 0.1 deg in PA and inclination.
    The deprojection with PA$=-15.4^\circ$ and Inc$=51.1^\circ$ minimizes the RMS (red box).    
    (c) The white solid line denotes the major and minor axes of the disk.
}
\label{fig:min}
\end{figure*}

\begin{figure*}[t!]
\begin{center}
\includegraphics[width=\textwidth]{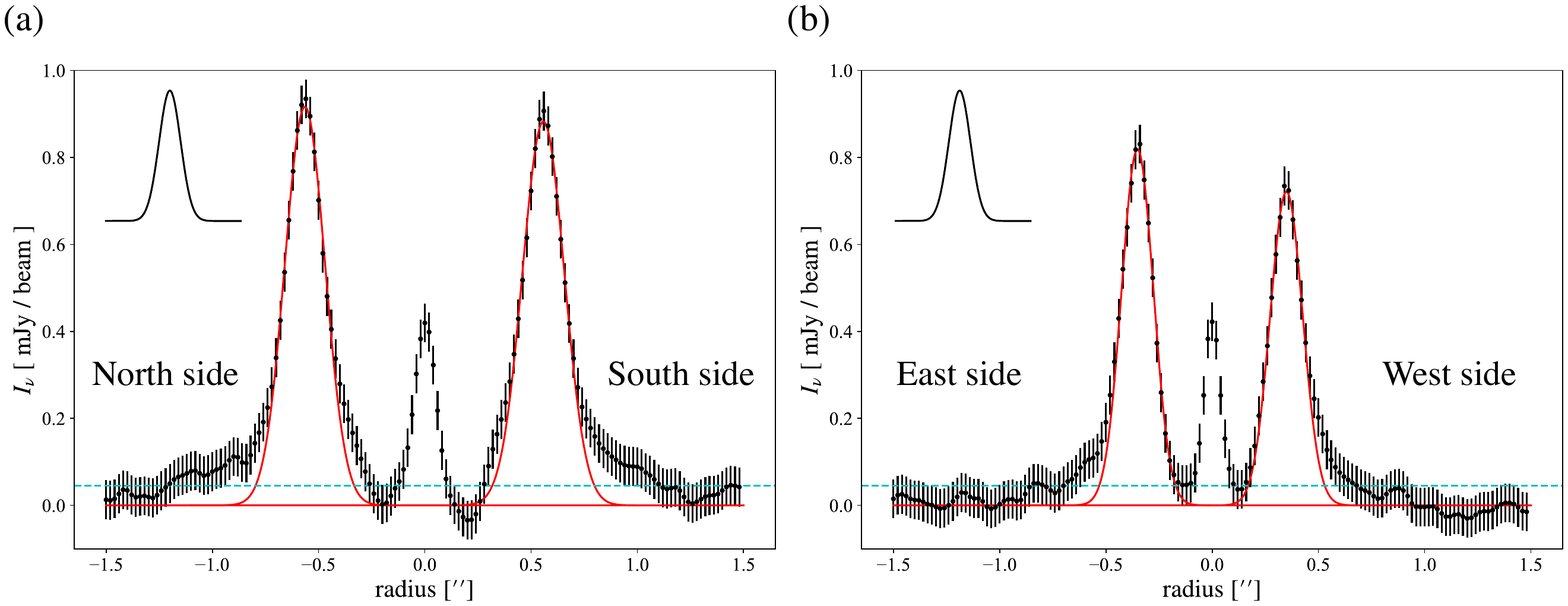}%
\end{center}
\caption{Radial profiles of the intensity along the (a) major and (b) minor axes (\figref{dust_continuum}b).
        The red line shows the best-fit curve in the Gaussian fitting at each peak.
        The error bar indicates the $3\sigma$, where the $1\sigma$ level is the RMS noise ($=14.9\ \rm{\mu Jy\ beam^{-1}}$).
        The black curve in the top left indicates the profile of the synthesized beam.
        The cyan dashed horizontal line is $3\sigma$.
        }
\label{fig:radpro}
\end{figure*}

  \begin{table}[t!]
  \tbl{Results of gaussian fitting at the major axis}{%
    \begin{tabular}{l|c|c} 
    \hline
      Parameter&North&South\\
      \hline
      $r_{peak}[\rm{}arcsec,\ au]$&
      $-$0.566, $-$102.3&0.559, 101.0\\
      $\rm{FWHM[arcsec,\ au]}$&
      0.216, 39.0&0.225, 40.7\\
      $I_\nu[\rm{mJy/beam}]$&
      0.922&0.889\\
      \hline
      \end{tabular}}\label{tab:radpromaj}
      \end{table}

      \begin{table}[t!]
      \tbl{Results of gaussian fitting at the minor axis}{%
        \begin{tabular}{l|c|c} 
        \hline
          Parameter&East&West\\
          \hline
          $r_{peak}[\rm{}arcsec,\ au]$&
          $-$0.351, $-$63.4&0.349, 63.1\\
          $\rm{FWHM[arcsec,\ au]}$&
          0.178, 32.2&0.183, 33.1\\
          $I_\nu[\rm{mJy/beam}]$&
          0.815&0.717\\
          \hline
          \end{tabular}}\label{tab:radpromin}
          \end{table}

The dust continuum image obtained by MS-CLEAN is shown in \figref{dust_continuum}.
The continuum emission has a total flux density of 55.9 mJy that was measured above $3\sigma$ in the emission.
The emission consists of a ring and a central source of 0.4~mJy.
The 2D Gaussian fitting of the point source on the image plane using \texttt{emcee} (\cite{foreman2013}) shows that this size is $42\pm 0.96$ mas $\times$ $25\pm 0.37$ mas and is comparable in size to the beam, thus the central source is unresolved.
This size can be regarded as the upper limit on the diameter of the inner dust disk. 

To characterize the ring structure, the center and tilt of the disk were determined.
We consider that the ring center is not necessarily located at the position of the central point source. 
Even when the physical center of the dust ring is located at the stellar position (and also the central component), the emergent emission from the ring can show asymmetry due to near-far asymmetry along the minor axis.
Therefore, we determined the center of the dust ring using the following methods.
First, we removed the emission corresponding to the unresolved inner disk at the center of the image by subtracting 0.4~mJy from the real part of all of the observed visibility. 
Secondly, we searched for the center of the ring structure.
This was performed in the visibility domain by shifting the phase center and minimizing the imaginary part of the concatenated visibility; in this process, all visibilities with any spatial frequencies were used.
\Figref{min}a shows the RMS level in each map synthesized with only the imaginary part as a function of the shift of phase center.
The RMS values of the images synthesized with only the imaginary part of the visibility were measured in the central region with a radius of 2 arcsec.
The minimum value of RMS was obtained for the phase center shift $\rm \Delta R.A. = -3$ mas and $\rm \Delta Dec. = 3$ mas.
Finally, we searched for the tilt of the disk.
This was performed by minimizing the RMS of the difference between the image and its $90^\circ$ rotation.
If the image is completely deprojected to face-on, the difference should be minimized.

\Figref{min}b shows the RMS level in each map of
difference between the image and its $90^\circ$-rotation as a function of PA and Inclination.
The result presented in this paper was obtained when the image was rotated 90$^\circ$ to the east, but the same result was obtained when the image was rotated to the west.
The minimum value of RMS was obtained for the deprojection with PA$=-15.4^\circ$ and Inc$=51.1^\circ$.

The surface brightness profiles along the major and minor axes are presented in \figref{radpro}.
We fit these profiles with three Gaussian components.
Two of them correspond to the ring, and the remaining corresponds to the central emission.  
The results of the Gaussian fitting along the major and minor axes are shown in tables \tabr{radpromaj} and \tabr{radpromin}, respectively.
The FWHM of the outer ring is about twice the beam of the dust continuum image.
We therefore consider that the width of the outer ring is marginally resolved.
We found that the radial profiles at the outer radii ($r\gtrsim 0.5$ arcsec) do not follow the Gaussian distribution.
According to the profile along the major axis, there is a weak emission ($>3\sigma$) that extends to $\sim 1.2$~arcsec from the central star.
More detailed analyses of the brightness distribution of dust continuum emissions in the visibility domain are presented in section~3.2.

In the radial profile along the minor axis, the peak on the east side is $\sim 10\%$ higher than that on the west side. 
This difference corresponds to $6.6\sigma$ (1$\sigma = 14.9\ \rm{}\mu Jy/beam$); thus, it is statistically significant.
The ring emission may be asymmetric along the minor axis, probably because the eastern side is the far side of the disk. 
The inner wall of the ring, which has a slightly higher temperature than the rest of the ring due to stellar irradiation, is observed in the eastern side (see section~4 for further discussion).


\subsubsection{CO line} 

\begin{figure*}[t!]
\begin{center}   
\includegraphics[width=\textwidth]{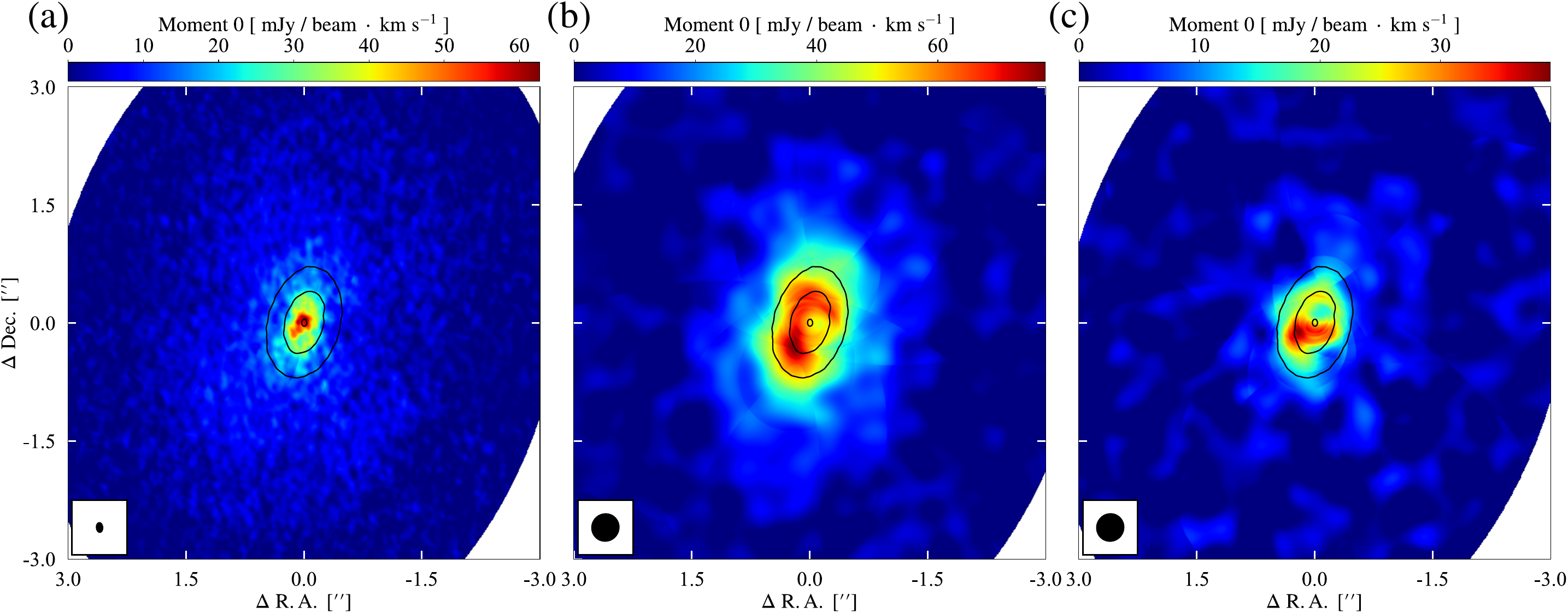}%
\end{center}
\caption{Moment 0 maps of (a)$\rm{}^{12}CO$(2$-$1), (b) $\rm{}^{13}CO$(2$-$1), (c) $\rm{}C^{18}O$(2$-$1).
  The black ellipse at the bottom left of each panel represents the beam size.
    The contours in these maps indicate the $20\sigma$ level of the dust continuum, where $1\sigma = 14.9\ \rm{}\mu Jy/beam$.
}
\label{fig:mom0}
\end{figure*}

\begin{figure*}[t!]
\begin{center}   
\includegraphics[width=\textwidth]{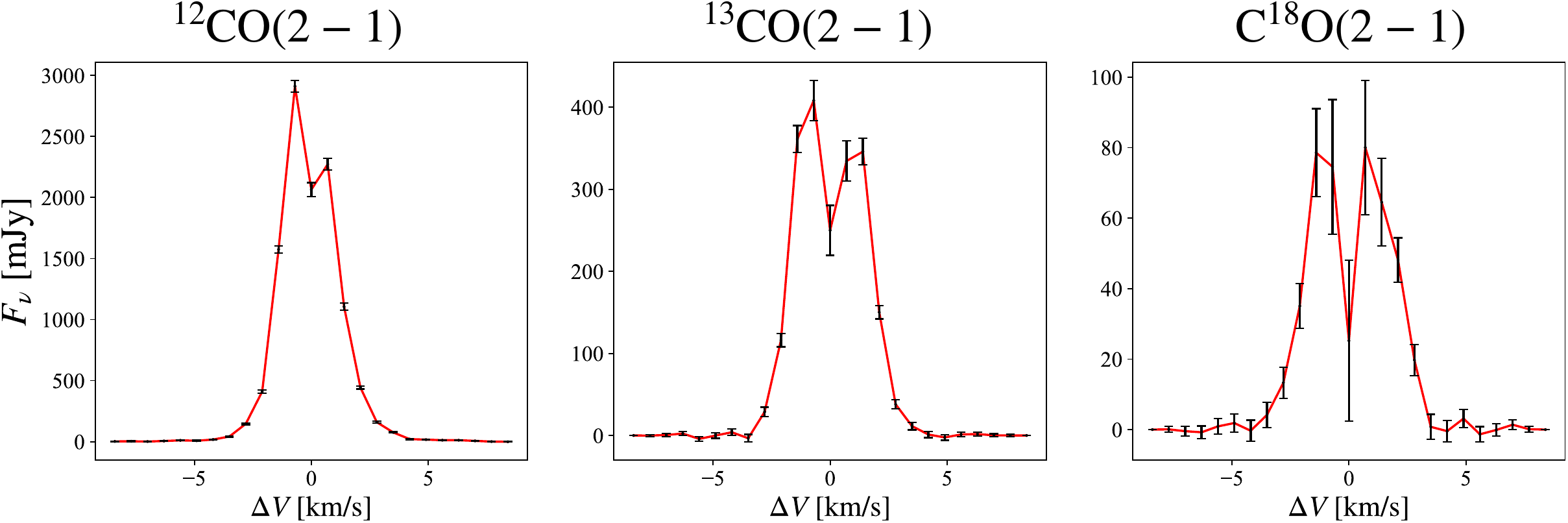}%
\end{center}
\caption{$\rm{}^{12}CO$(2$-$1), $\rm{}^{13}CO$(2$-$1), and $\rm{}C^{18}O$(2$-$1) emission line spectra.
The error bars were obtained by error propagation from the image RMS.
}
\label{fig:spectrum}
\end{figure*}

\begin{figure*}[t!]
\begin{center}
\includegraphics[width=\textwidth]{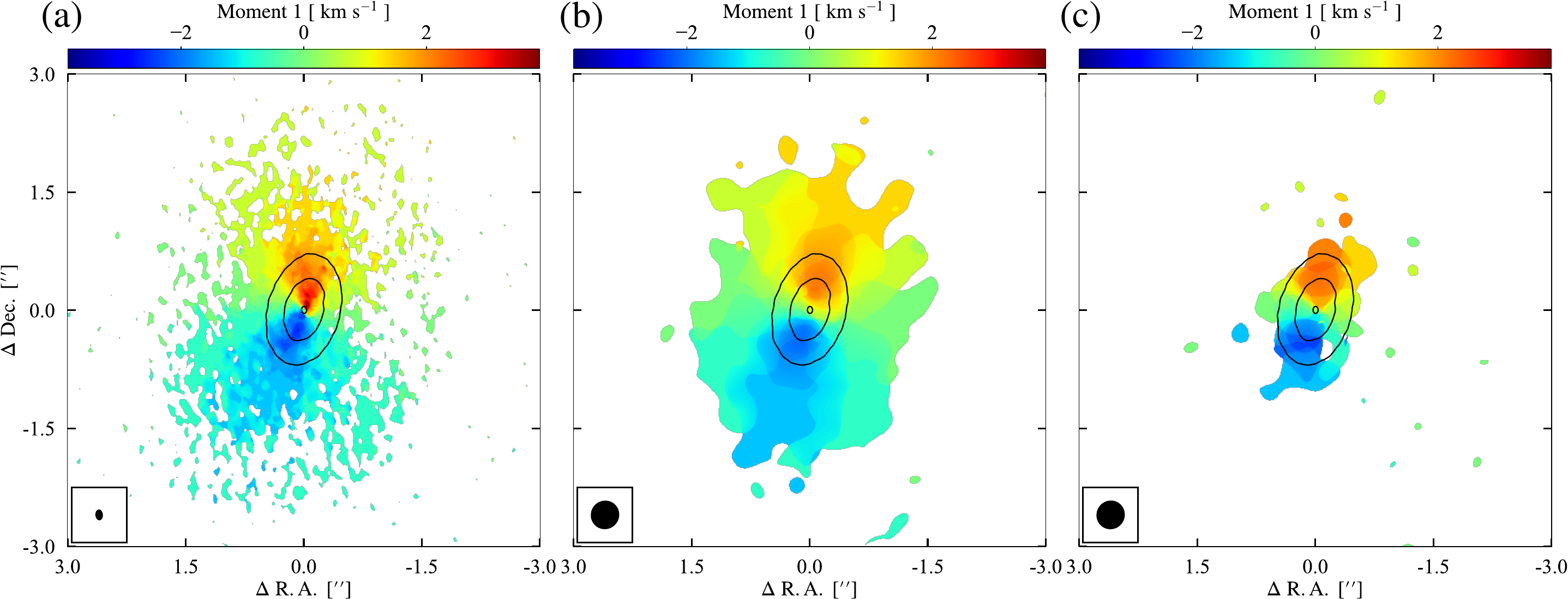}%
\end{center}
\caption{Moment1 maps of (a) $\rm{}^{12}CO$(2$-$1), (b) $\rm{}^{13}CO$(2$-$1), (c) $\rm{}C^{18}O$(2$-$1).
  The black ellipse at the bottom left of each panel represents the beam size. 
    These color bars show the velocities based on systemic velocity $V_{\rm{LSRK}}=4.1\ \rm{km/s}$. 
    The contours in these maps indicate the $20\sigma$ level of the dust continuum, where $1\sigma = 14.9\ \rm{}\mu Jy/beam$.
    When creating the moment 1 maps, the clipping levels were set in 3$\sigma$.
}
\label{fig:mom1}
\end{figure*}

\begin{figure*}[t!]
\begin{center}
\includegraphics[width=\textwidth]{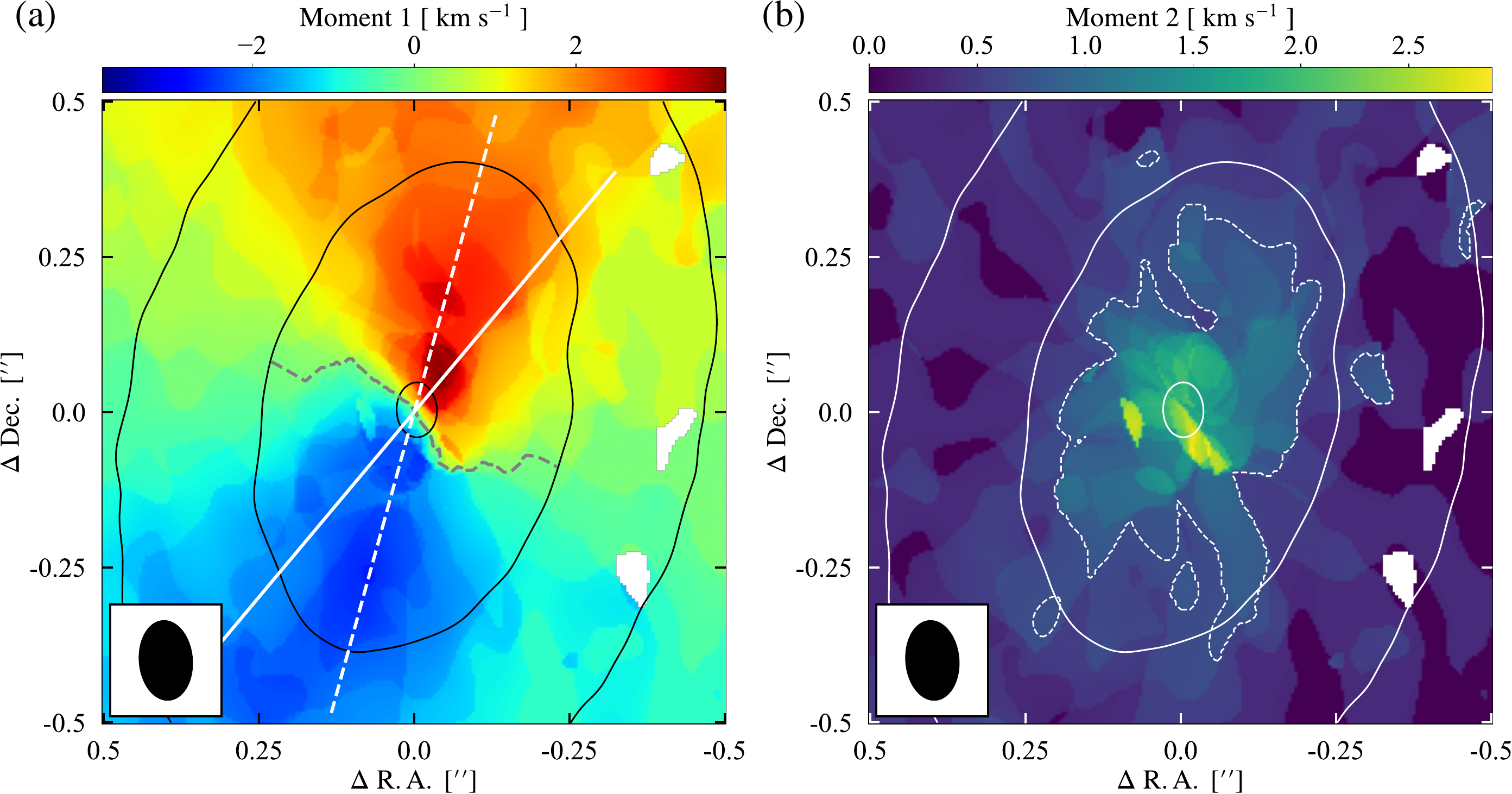}%
\end{center}
\caption{
    (a) Zoom-up moment 1 map made from the $\rm{}^{12}CO$(2$-$1) emission.
    The black ellipse at the bottom left of each panel represents the beam size.
    The gray dashed contour indicates the systemic velocity.
    The dashed and solid lines denote the major axis of the outer ring and an inner disk (see the discussion in section 4.2), respectively.
    (b) Moment 2 map of $\rm{}^{12}CO$(2$-$1).
    When creating the moment 2 map, the clipping levels were set at 3$\sigma$.
    The synthesized beam size is shown at the bottom left.
    The dashed contours indicate the velocity dispersion of 0.7 km/s.
    The solid contours in these maps indicate the 20$\sigma$ level of the dust continuum, where $1\sigma = 14.9\ \rm{}\mu Jy/beam$.
}
\label{fig:mom1_zoom}
\end{figure*}

\begin{figure*}[t!]
\begin{center}
\includegraphics[width=\textwidth]{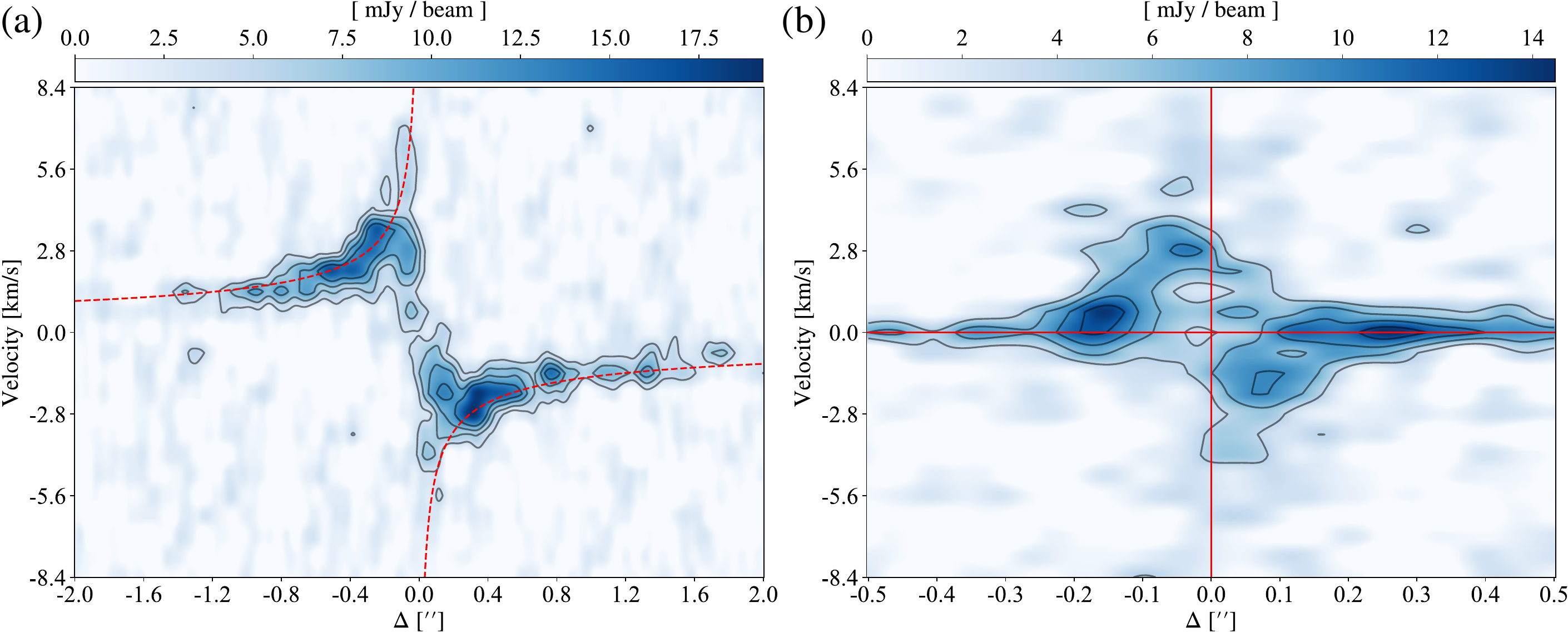}%
\end{center}
\caption{PV diagram made from the $\rm{}^{12}CO$(2$-$1) cube along the (a) major and (b) minor axes of the outer ring.
      Contours are shown at [3,5,7,9]$\sigma$, where the $1\sigma$ level is $1.34\ \rm{mJy/beam}$.
      (a): The red curves show the Keplerian velocity curve for the central star of 0.78 $M_{\odot}$, and an outer disk inclination of 51.1$^\circ$.
      (b): The solid red lines denote the center position, and the systemic velocity of 0.
      There exists a velocity gradient in the vicinity of the center.
    }
\label{fig:pv}
\end{figure*}

\begin{figure*}[t!]
\begin{center}
\includegraphics[width=\textwidth]{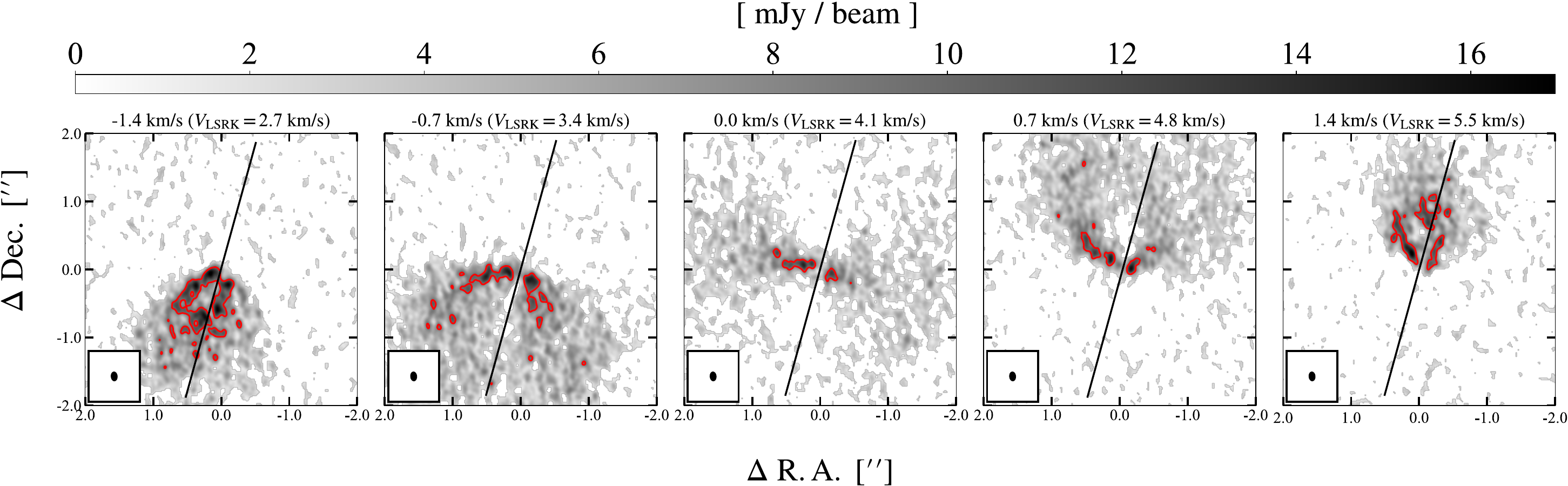}%
\end{center}
\caption{
The butterfly patterns on panels of $\Delta V = -1.4-1.4 \rm{}km/s$.
The red contours indicate the 10$\sigma$ level of the $^{12}$CO$(2-1)$, where $1\sigma = 1.34\ \rm{}mJy/beam$.
The black lines denote the major axis of the disk.
}
\label{fig:twin}
\end{figure*}

The moment 0 (integrated intensity) maps of the CO line emission are shown in \figref{mom0}.
Each panel corresponds to $^{12}$CO data, $^{13}$CO data, and C$^{18}$O data. 
We summarized the integrated flux of each moment 0 map in \tabref{imageparam}.
The $^{12}$CO line shows that the line emission comes from the elliptical region whose major axis well matches with that of the dust ring. 
We observe the $^{12}$CO emission inside the dust ring, which peaks at the location where the dust emission at the center of the ring is observed.
The moment 0 map of the $^{13}$CO line emission also shows an extended distribution along the major axis of the dust ring.
Contrary to the $^{12}$CO emission, the $^{13}$CO emission is concentrated in the inner edge of the dust ring, and also shows an east-west asymmetry, as in the case of the dust continuum.
This suggest that the disk is optically thinner for the $^{13}$CO emission, tracing both column density and temperature, and the peak of the image seems to delineate a high gas density near the dust ring.
The distribution of C$^{18}$O emission resembles that of $^{13}$CO, but the signal is less significant; the intensity of the C$^{18}$O emission in channel maps has at most  10.6$\sigma$ with 0.35$\arcsec$ beam, which is 2.3 times lower than that of the $^{13}$CO emission.

As shown in \figref{spectrum}, all line emissions indicate the spectra in which the blue channels are brighter than the red channels.
In the case of absorption by the foreground cloud, it is relatively optically thick for the $^{12}$CO line, resulting in a relatively high extinction of the $^{12}$CO line spectra.
However, since all the emission lines show similar spectra, this is most likely a result of the gas distribution in the disk rather than absorption by the foreground cloud.

\Figref{mom1} shows the moment 1 (intensity-weighted velocity) map of CO lines.  
The velocity clearly shows the rotation pattern, and the direction of the large velocity gradient coincides with that of the major axis of the dust ring.
The velocity structure revealed in the moment 1 maps of both $^{13}$CO and C$^{18}$O agree with that of $^{12}$CO, which is consistent with rotation in a disk.

However, we see a peculiar velocity pattern in the vicinity of the central star. 
\Figref{mom1_zoom}a is the moment 1 map of $^{12}$CO emission near the central star. 
Although the line-of-sight velocity should be zero along the minor axis of a coplanar disk, the velocity pattern inside the dust ring is distorted.

No significant difference is found between the high velocity dispersion position (above 2~km/s) in the moment~2 map (velocity dispersion) shown in \figref{mom1_zoom}b and the point source position in the range of the beam size.
We consider that the central dust component is cospatial with the central star, but the spatial resolution may be insufficient to discuss the overlap between the velocity dispersion map and the central dust component. 
The high velocity dispersion ($\approx 3 \rm{}km/s$) along the minor axis of the disk is seen from the moment~2 map, but this component is derived from a low SN emission ($< 4\sigma$); whether this is real is unclear.

\Figref{pv}a shows the position-velocity diagram (PV diagram) of $^{12}$CO emission along the major axis of the ring.  
Assuming a stellar mass of 0.78$M_{\odot}$ (\cite{manara2017}) and a disk inclination of 51.1$^\circ$, as derived from the ellipse fitting to the continuum emission (section~3.1.1), the Keplerian rotation curves are shown by red curves in \figref{pv}a.
The velocity field of the $^{12}$CO emission is almost consistent with the Keplerian motion of the gas.
However, there is another velocity gradient in the PV diagram along the minor axis (\figref{pv}b) that cannot be accounted for by the pure Keplerian motion in the disk, and there are high-velocity components with a velocity shift of 2$–$7 km/s from the systemic velocity in the vicinity ($<$0.1~arcsec) of the central star.
The distortion of the velocity pattern observed in moment 1 map may be caused by a warp (\cite{casassus2015}; \cite{mayama2018}) of the inner disk or fast radial flow (\cite{rosenfeld2014}).
We discuss the nature of the velocity pattern in detail in section~4.

As shown in figures~\figr{chan_12CO}$-$\figr{chan_C18O} of appendix 1, we obtained the channel maps of each CO isotopologue emission.
\Figref{twin} shows the butterfly patterns in $\Delta V = -1.4-1.4\ \rm{}km/s$ as red contours, corresponding to the 10$\sigma$ level of the $^{12}$CO$(2-1)$ ($1\sigma = 1.34\ \rm{}mJy/beam$).
As mentioned in section~3.1.1, the east side corresponds to the far side, where the butterfly pattern can be identified more clearly, when the emissions from both upper and lower disk surfaces are observed (\cite{gregorio2013}; \cite{tsukagoshi2019}).

\begin{figure*}[t!] 
\begin{center}
\includegraphics[width=\textwidth]{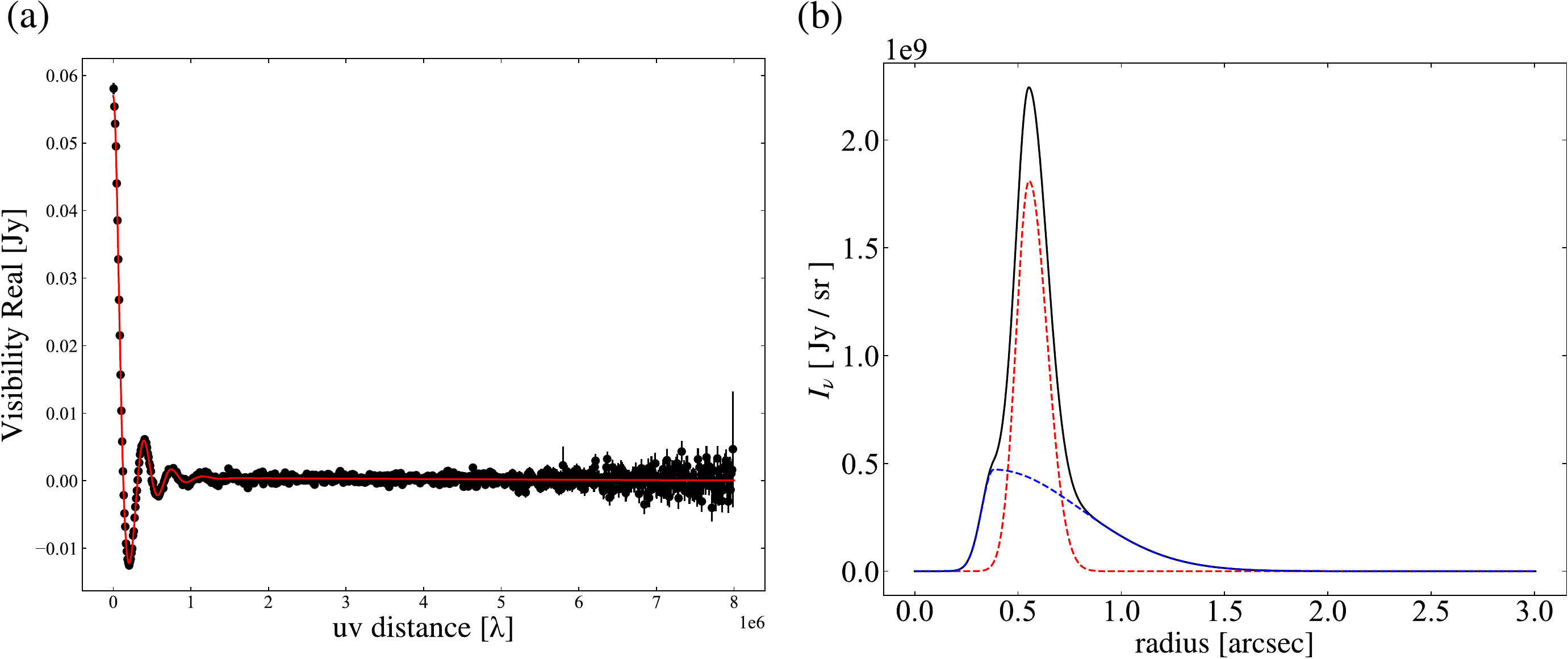}%
\end{center}
\caption{
  (a) Real part of the deprojected visibility for the observations (black dots) and best-fit model (red curve).
    (b) Radial profile of intensity as Fourier transform of the best-fit visibility (black line).
    The model consisted of two Gaussian ring components (blue and red dashed lines). 
}
\label{fig:visfit}
\end{figure*}

\begin{figure*}[t!]
\begin{center}
\includegraphics[width=\textwidth]{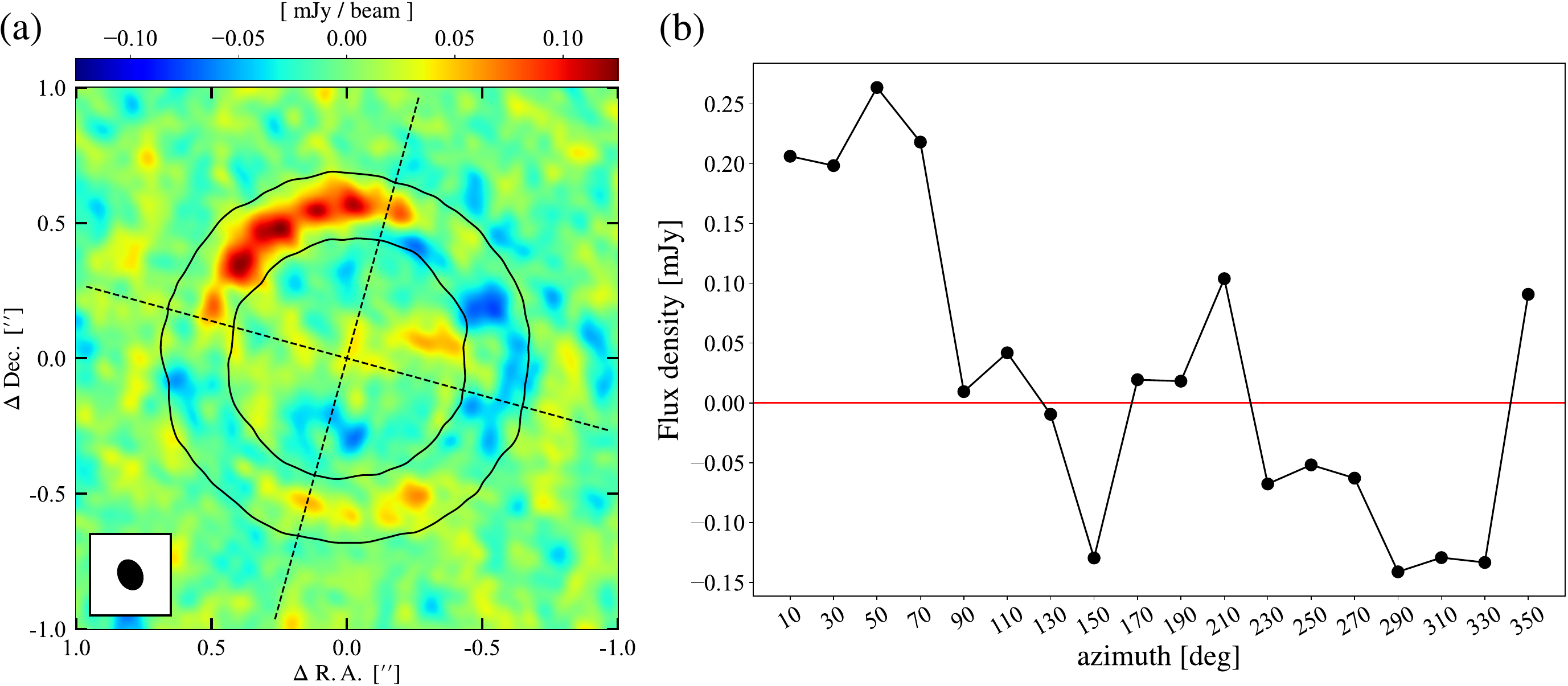}%
\end{center}
\caption{(a): Residual map obtained by subtracting the model visibility from the deprojected visibility.
  The synthesized beam ($0.114^{\prime\prime} \times 0.088^{\prime\prime}$) is shown at the bottom left.
    The white contours denote the $20\sigma$, where the $1\sigma$ level is the RMS noise (= $13.8\ \rm{\mu Jy\ beam^{-1}}$).
    From the figure, there is a quadrupole pattern of asymmetric intensity on the outer ring.
    (b): Azimuth profile of flux density measured in the range of $\Delta$PA = 20$^\circ$ and the radius of 0.3$-$1 arcsec in the residual map.
}
\label{fig:visres}
\end{figure*}


\subsection{Analysis of dust continuum emission in visibility domain}

In this section, we investigate the asymmetry of the dust emission rings in the deprojected visibility domain.
The Fourier transform of the azimuthally symmetric distribution becomes a Hankel transform (\cite{Baddour_2009})
\begin{eqnarray}
V_{\rm{}real}(u)= 2\pi \int_{0}^{\infty}I(r)J_0(2\pi u r )rdr,
\end{eqnarray}
where $V_{\rm{real}}(u)$, $I(r)$, and $J_{0}$ are the visibility real part of the deprojected uv-distance $u$, intensity of the distance from the center $r$, and zeroth-order Bessel function of the first type, respectively.

To fit the bright ridge of the ring as well as the extended periphery, we adopted a function composed of two radially asymmetric Gaussians as the model for radial brightness distribution, as follows:
\begin{equation}
    I(r)=\sum_{i=1}^2I_{{\rm{}r}i, \rm{}peak}\exp\left[-\ln 2\frac{(r-R_{{\rm{}r}i})^2}{{\rm{}HWHM}_{{\rm{}r}i}^2}\right],
\end{equation}
\begin{equation}
{\rm{}HWHM}_{{\rm{}r}i}=
\left\{
\begin{array}{ll}
    {\rm{}HWHM}_{{\rm{}in}, {\rm{}r}i} & (r < R_{{\rm{}r}i}),\\
    {\rm{}HWHM}_{{\rm{}out}, {\rm{}r}i} & (r \geq R_{{\rm{}r}i}),
\end{array}
\right.
\end{equation}
\begin{equation}
      I_{{\rm{}r}i, \rm{}peak}=F_{{\rm{}tot,r}i}\left[2\pi \int_{0}^{\infty}\exp\left[-\ln 2\frac{(r-R_{{\rm{}r}i})^2}{{\rm{}HWHM}_{{\rm{}r}i}^2}\right]J_0(0)rdr\right]^{-1},
\end{equation}
where $F_{{\rm{}tot,r}i}$ is the total flux density of of the $i$-th ring.
We determine eight parameters ($R_{\rm{r1}}$, $\rm{HWHM_{in,r1}}$, $\rm{HWHM_{out,r1}}$, $F_{\rm{}tot,r1}$, $R_{\rm{r2}}$, $\rm{HWHM_{in,r2}}$, $\rm{HWHM_{out,r2}}$, $F_{\rm{}tot,r2}$) that best fit the observed visibility profile by using \texttt{emcee} (\cite{foreman2013}).  
The best-fit parameters are summarized in \tabref{visfit3} and the visibility profile for the best-fit model is shown in \figref{visfit}a. 
\Figref{visfit}b presents the radial brightness distribution expressed by equation~(2).

\Figref{visres}a shows the image substracting the symmetric component from the original image.
The residual asymmetric components may be characterized by a quadrupole pattern: positive residuals in the northeast and southwest to the star and negative residuals in the southeast and northwest.  
\Figref{visres}b shows the azimuth profile of flux density measured in the range of $\Delta$PA =  $20^\circ$ and the radius of 0.3$-$1 arcsec in the residual map.
This pattern shows east-west asymmetry, in which the east side appears brighter than the west side on the whole.

\begin{table}[t]
\tbl{Fitting parameters of two-gaussian model}{%
  \begin{tabular}{l|c} 
  \hline
    Parameter&Value\\
    \hline
    $R_{\rm{r1}}\ [\rm{}arcsec, au]$&
    0.553, 100\\
    $\rm{HWHM_{in,r1}\ [arcsec, au]}$&
    0.075, 13.5\\
    $\rm{HWHM_{out,r1}\ [arcsec, au]}$&
    0.101, 18.3\\
    $F_{\rm{}tot,r1}\ \rm{}[mJy]$&
    28.5\\
    $R_{\rm{}{r2}}\ [\rm{}arcsec, au]$&
    0.385, 69.5\\
    $\rm{HWHM_{in,r2}\ [arcsec, au]}$&
    0.071, 12.9\\
    $\rm{HWHM_{out,r2}\ [arcsec, au]}$&
    0.498, 89.9\\
    $F_{\rm{}tot,r2}\ [\rm{}Jy]$&
    28.0\\
    \hline
    reduced $\chi^2$&
    1.14\\
    \hline
    \end{tabular}}\label{tab:visfit3} 
    \end{table}

\subsection{Physical properties of the disk }

\begin{figure*}[t!]
\begin{center}
\includegraphics[width=\textwidth]{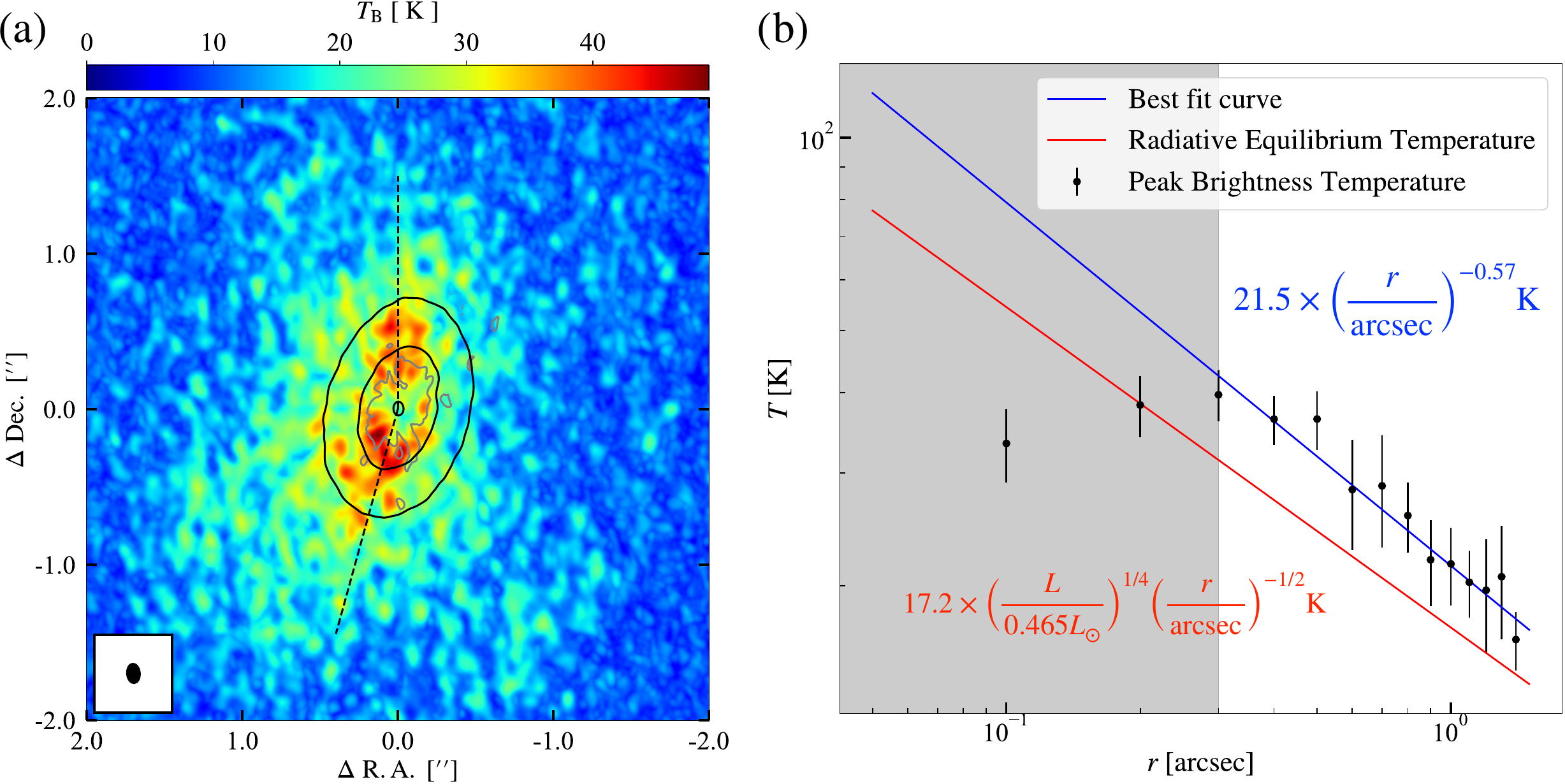}%
\end{center}
\caption{(a): Peak brightness temperature of $\rm{}^{12}CO$(2$-$1) emission.
  The synthesized beam size is shown at the bottom left.
    The black contours indicate the 20$\sigma$ level of the dust continuum, where $1\sigma = 14.9\ \rm{}\mu Jy/beam$.
    The gray contours indicate the velocity dispersion of 0.7 km/s.
    (b) Radial profile of the straight line in (a).
    The blue line indicates the curve fitted in area other than gray.
    The red line indicates the radiative equilibrium temperature of a blackbody.
}
\label{fig:mom8}
\end{figure*}

\begin{figure}[t!]
\begin{center}
\includegraphics[width=.48\textwidth]{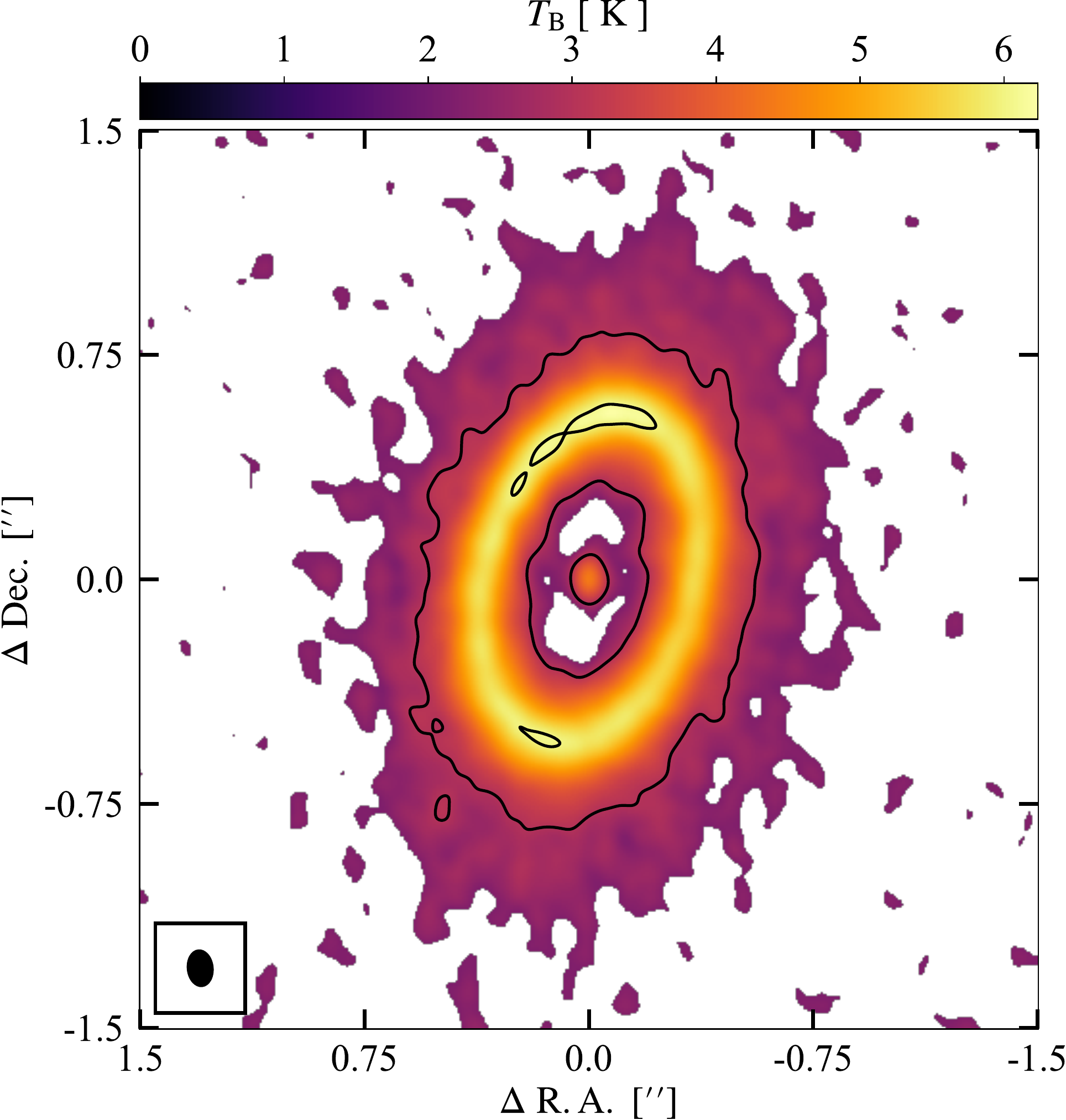}%
\end{center}
\caption{Brightness temperature of the dust.
    The black ellipse at the bottom left shows the synthesized beam of the dust continuum image.
    The contour indicates 3 K and 6 K.
    The area where the emission is shown to be less than 3$\sigma$ in \figref{dust_continuum}a is masked, where $1\sigma = 14.9\ \rm{}\mu Jy/beam$.
    The brightness temperature indicates the lower limit of the dust temperature.
}
\label{fig:dust_K}
\end{figure}

\begin{figure*}[t!]
\begin{center}
\includegraphics[width=\textwidth]{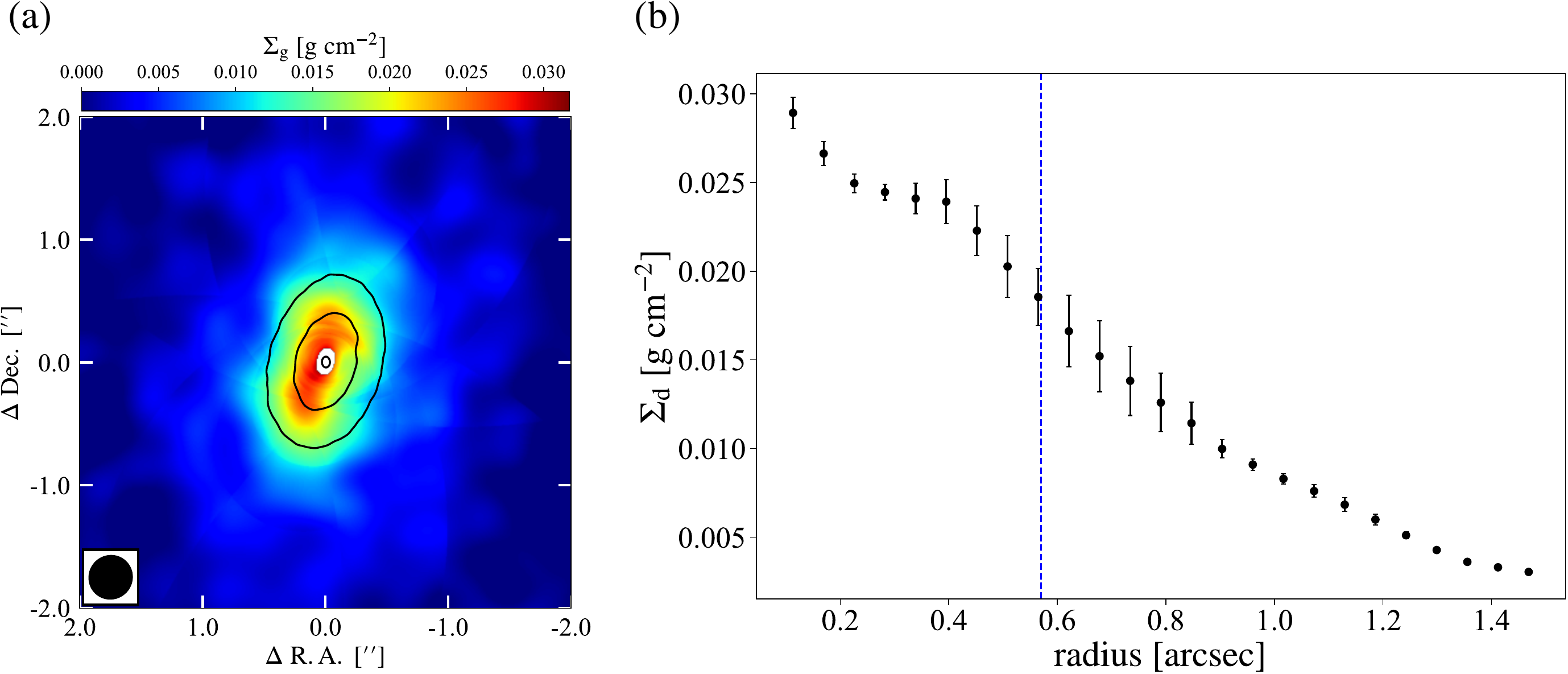}%
\end{center}
\caption{(a): Gas column density $\Sigma_{\rm{g}}$ under the assumptions of local thermal equilibrium (LTE) and the temperature distribution expressed by equation~(5).
  The synthesized beam size is shown at the bottom left. 
    (b) Radial profile of the gas column density along the major axis in (a), which was combined with each side.
    The error bar is the standard deviation of the mean within the range of $\pm10$ mas.
    The blue dashed lines indicate the peak position of the dust ring.
}
\label{fig:sigma_g}
\end{figure*}

\subsubsection{Dust Column Density}

From the continuum emission, we estimated the optical depth and column density of dust in the SY Cha system.
Here, we show simple estimates based on observed quantities because we only have Band 6 data.
We estimate the optical depth of the dust ring $\tau_{\rm{}d}$ by
\begin{equation}
I_{\nu} = B_{\nu}(T) (1-\exp (-\tau_{\rm{}d})),
\end{equation}
where $I_{\nu}$ is the surface brightness of the continuum, and $B_{\nu}$ is the Planck function. 
As shown in \figref{mom8}, we adopt a temperature distribution derived from the power-law fit to the peak brightness temperature of the $\rm{}^{12}CO$(2$-$1), as 
\begin{eqnarray}
T = 415.8\ \Big(\frac{r}{\rm{au}}\Big)^{-0.57}\ \rm{K}.
\end{eqnarray}
Here, we fitted the peak brightness temperature only in the region where the velocity dispersion of the $^{12}$CO cube is less than 0.7 km/s (i.e., the channel width).
The temperature near the disk mid-plane, where most of the dust particles are distributed, is expected to be lower than that in the upper layer of the disk traced by the $\rm{}^{12}CO$ line emission (\cite{Chiang1997}).
Therefore, the temperature expressed in equation~(6) is the upper limit of the disk temperature.
This temperature is 1.8 times higher than the radiative equilibrium temperature for a blackbody, which is as follows in the case of SY Cha, 
\begin{eqnarray}
T_{\rm{rt}} = 231.22\ \Big(\frac{L}{0.465L_{\odot}}\Big)^{\frac{1}{4}}\Big(\frac{r}{\rm{}au}\Big)^{-\frac{1}{2}}\ \rm{K}, \label{ret}
\end{eqnarray}
where $0.465 L_{\odot}$ is the luminosity of SY Cha (\tabref{sycha}).

The brightness temperature of the dust emission (\figref{dust_K}) can be lower than that of equation (5), due to beam dilution effects and dust scattering (\cite{baobab2019}).
The dust scattering depends on the dust compositions (\cite{tazaki2016}) and makes the dust emission dimmer (\cite{soon2019}; \cite{ueda2020}).
However, high-spatial-resolution data for SY Cha have been taken only in Band 6 presented in this paper. 
Thus, we have ignored these effects and estimated the lower limit of the optical depth of the dust ring $\tau_d$ at 225~GHz.

Assuming that the opacity of unit dust mass is
\begin{eqnarray}
\kappa_{\rm{d}}=10\Big(\frac{\nu}{10^{12}\rm{Hz}}\Big)^\beta\ \rm{cm^2\ g^{-1}}
\end{eqnarray}
(\cite{beckwith1990}), we estimated the dust column density $\Sigma_{\rm d}$($=\tau_{\rm d}/\kappa_{\rm d}$). 
At 225~GHz, $\kappa_{\rm d}$ derived from equation~(8) is 2.25~cm$^2$/g under the assumption that $\beta$ = 1.
The total dust mass integrated over the entire disk was estimated to be $\sim 1.0 \times 10^{-4}~M_{\odot}$.
This value may be regarded as the lower limit of the total dust because the assumed temperature is the upper limit of the disk temperature.

The lower limit of the total dust mass of the central compact ring is $\approx$0.05 $M_{\oplus}$ ($\approx 1.7\times10^{-7}$ $M_{\odot}$).
This suggests that the inner ring might have a similar mass and orbit to Mercury in our solar system.
In addition, there is no remarkable excess emission at $\lambda \lesssim 3\mu \rm{}m$ in the SED of SY Cha, as described in section~1. 
This indicates that the disk does not harbor hot ($\gtrsim 1000$~K) dust particles within at least $\sim 0.1$~au from the central star (see equation~(5)).  
Although the central point source is not spatially resolved even with the highest-spatial-resolution image, we consider that the central compact source has a ring-like structure as indicated in the case of DM~Tau (\cite{Kudo2018}).  
The upper limit of the size of the central ring is $\sim 3.5$~au since the beam size is $\sim$7~au.

The pressure scale height $h$ is defined by $c_s/\Omega_{\rm{}K}$, where $c_s$ and $\Omega_{\rm{K}}$ are the sound speed and angular velocity of the Keplerian rotation, respectively.
The thickness of gas can be estimated as 2$h_{\rm{}d}$ = 25 au at a radius of 100~au when the temperature distribution is expressed by equation~(5).
Because the mm-size dust is sometimes concentrated near the disk mid-plane, this estimate provides the upper limit of the thickness of the dust ring.  
However, if the dust particles are well mixed with gas, the thickness is comparable to the ring radial width of 32~au (\tabref{visfit3}), suggesting that the outer ring has a torus-like shape.
Such geometry has also been observed in the Sz~91 system (\cite{tsukagoshi2019}).

\subsubsection{Gas column density}

We estimated the gas column density using the results of the $\rm{}^{13}CO$(2$-$1) emission, assuming local thermal equilibrium (LTE).  
We first derived the optical depth of the gas $\tau_{\rm g}$ for each channel map by
\begin{eqnarray}
I_{\rm{g}}=B_{\nu}(T_{\rm{g}})(1-\exp(-\tau_{\rm{g}})),
\end{eqnarray}
where $T_{\rm g}$ is the adopted gas temperature, expressed by equation~(5).
The optical depth $\tau_{\rm{}g}$ at each position of a channel map is integrated along the velocity axis, and then converted into a column density of $^{13}$CO, as described in detail in appendix 3. 
\Figref{sigma_g}a is the map of $\Sigma_{\rm{}g}$ under the assumption of the abundance ratio of $\rm{}^{13}CO/H_2 = 1/67 \times 10^{-4}$ (\cite{Qi2011}).
\Figref{sigma_g}b shows the radial profile of the gas column density along the major axis.  
The peak of the gas column density ($\approx0.4\ \rm{}arcsec$) may be located inside the dust ring.
The total gas mass is estimated to be $2.2\times 10^{-4}~M_{\odot}$.  
The total integrated flux of C$^{18}$O corresponds to the total gas mass of $\sim 3.7\times 10^{-4}~M_{\odot}$ assuming the C$^{18}$O abundance of C$^{18}$O$/$H$_2=1/444 \times 10^{-4}$ (\cite{Qi2011}), which is consistent with the gas mass estimated from $^{13}$CO emission within a factor of 2.  
The abundance of CO in protoplanetary disks may be lower than the corresponding interstellar medium (ISM) values by two orders of magnitude (\cite{Zhang2020}) and therefore the gas mass may be underestimated by a factor of 10$-$100.


\begin{figure*}[t]
\begin{center}
\includegraphics[width=\textwidth]{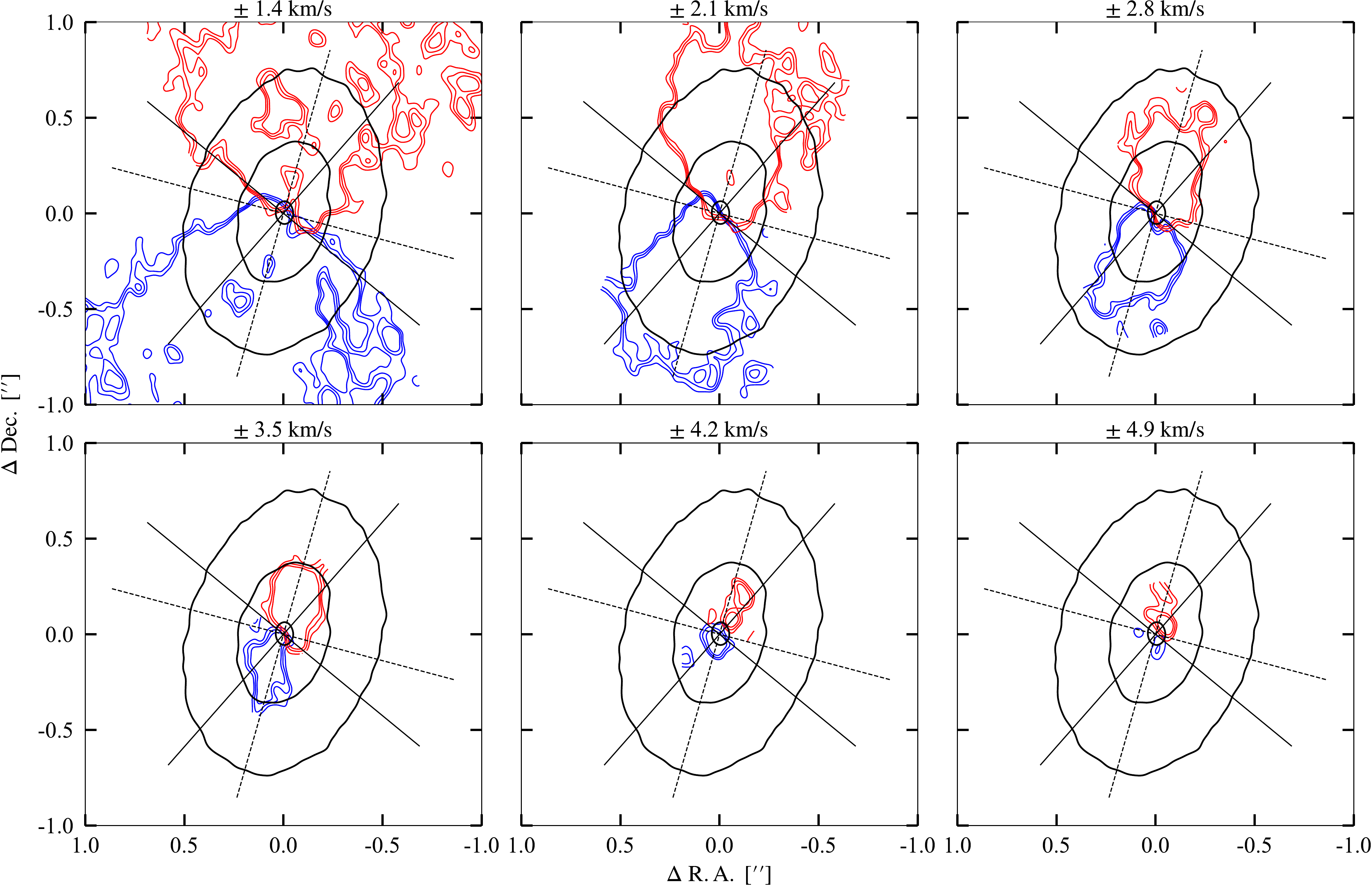}%
\end{center}
\caption{Channel map pf the $\rm{}^{12}CO$(2$-$1) emission imaged at a channel width of $0.7\ \rm{km/s}$.
  The velocities based on the systemic velocity ($V_{\rm{LSRK}}=4.1\ \rm{km/s}$) are labeled at the top of each plot.
    The red and blue contours denote the redshifted and blueshifted emissions, respectively. 
    The contour levels are $3\sim 9\sigma$, where the $1\sigma$ level is $1.34\ \rm{mJy/beam}$.
    The dashed lines represent the major and minor axes of the outer dust disk (PA = $-$15.4$^\circ$).
    The solid lines are the major and minor axes of the inner dust disk (PA = $-$40$^\circ$).
}
\label{fig:chanvel}
\end{figure*}

\begin{figure}
\begin{center}
\includegraphics[width=.45\textwidth]{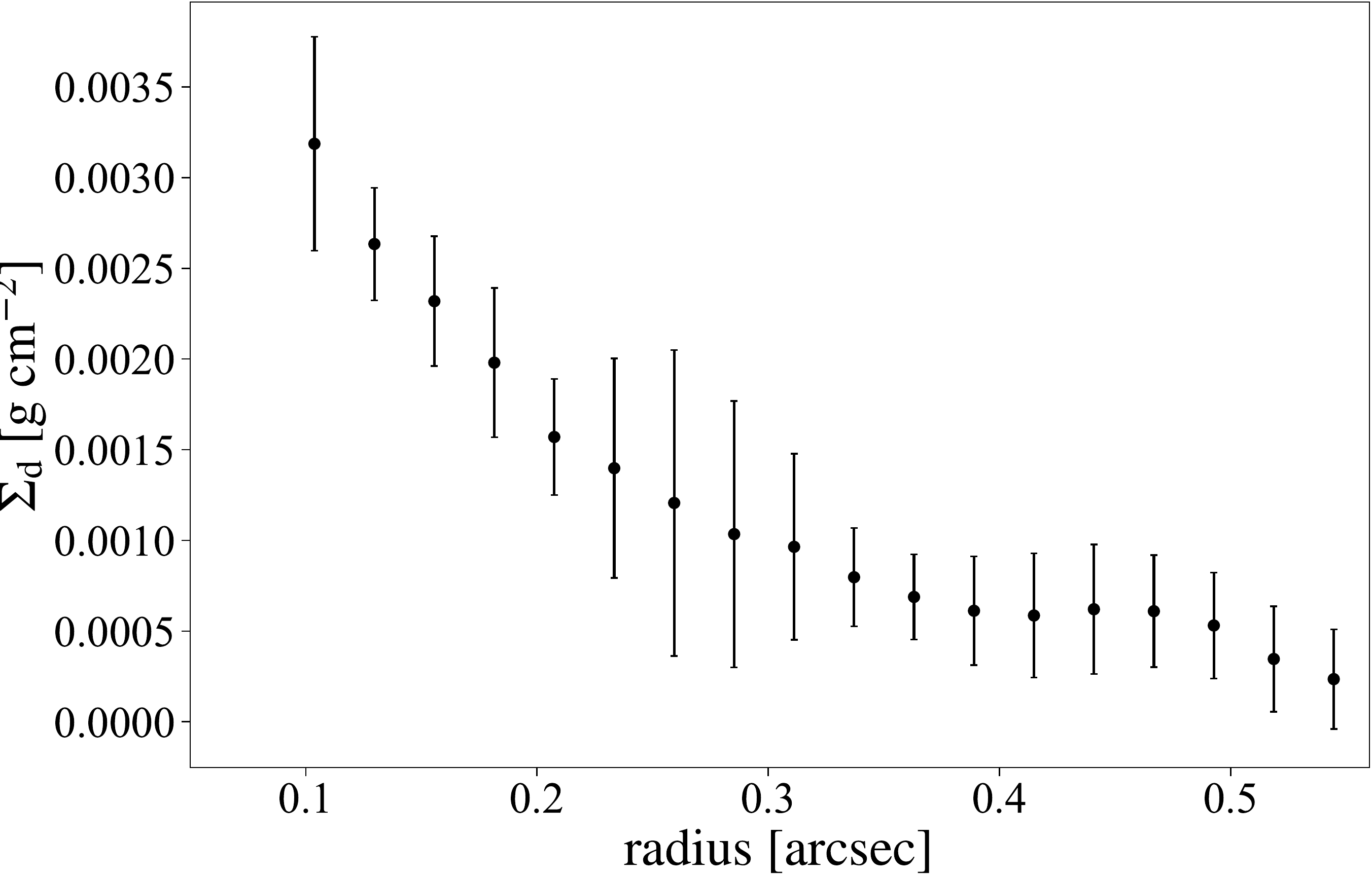}%
\end{center}
\caption{Radial profile of the gas column density calculated from $^{12}$CO along the minor axis, which was combined with each side.
  The error bar is the standard deviation of the mean within ±5 mas.}
  \label{fig:sigma12}
  \end{figure}

  \begin{figure*}[t]
  \begin{center}
  \includegraphics[width=\textwidth]{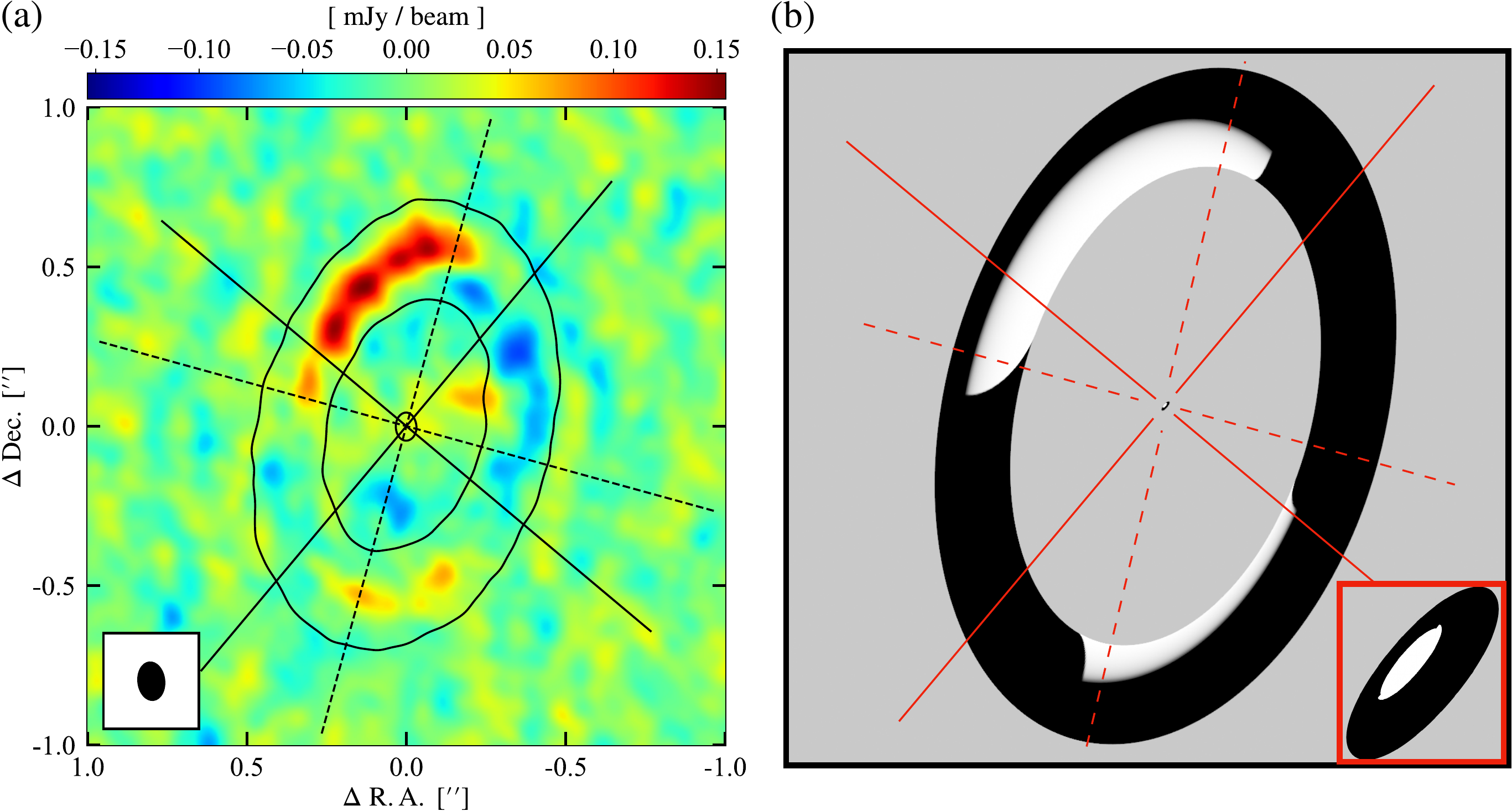}%
  \end{center}
  \caption{(a): Residual map projected \figref{visres}a.
    The synthesized beam size ($0.146^{\prime\prime} \times 0.110^{\prime\prime}$) is shown at the bottom left.
      The black contour indicates the $20\sigma$ level of the dust continuum, where $1\sigma = 14.9\ \rm{}\mu Jy/beam$.
      The dashed lines show the major and minor axes of the outer ring.
      The solid lines show the major and minor axes of the inner disk. 
      (b): Schematic diagram of the torus model that has an inner ring (PA=$-40^\circ$, inc.=$75^\circ$) and the outer ring (PA=$-15.4^\circ$, inc.=$51.1^\circ$), setting the radius ratio between the outer and the inner torus to 102:3, and setting the scale height of the two rings relative to their radius to 0.2.
           As with (a), the red lines show the major and minor axes of the inner ring shown at the bottom-right corner, and the red dashed lines show those of the outer ring.
  }
\label{fig:dust_model}
\end{figure*}

\section{Discussion} 

We detected the distortion of the line-of-site velocity pattern that cannot be explained by the Keplerian rotation of a geometrically thin disk.
As shown in \figref{mom1_zoom}a, the iso-velocity contour close to the center deviates from the minor axis of the ring by $\sim 25^\circ$. 
\Figref{chanvel} shows a part of the channel maps of $\rm{}^{12}CO$(2$-$1) emission.
Each panel compares the distribution of components blue-shifted and red-shifted by the same amount from the systemic velocity.
The distortion of the velocity pattern can be observed in all the panels up to $\Delta V=\pm$4.2 km/s, and the direction of the velocity gradient appears to change with velocity.

Other transitional disk systems show velocity distortion in the vicinity of the central star (e.g., \cite{casassus2015}; \cite{mayama2018}).
There are two possible mechanisms causing similar velocity distortion patterns: radial flow and warped inner disk (e.g., \cite{rosenfeld2014}).
Here, we discuss these two possibilities for the SY Cha system.

\subsection{Radial flow}
In this section, we discuss the possibility of radial flows in the disk.
In the vicinity of the central star, the line-of-site velocity is blue-shifted in the east and red-shifted in the west along the minor axis of the outer ring.  
Because the far side of the disk is in the east (see section~3), this distribution indicates that the velocity vector should be oriented towards the central star.  
Thus, fast infall may be happening in the vicinity of the central star.
As shown in the PV diagram along the minor axis of the outer disk (\figref{pv}b), we confirmed the non-zero velocity components within several tens of au.
The infall velocity $v_{\rm{}in}$ and the line-of-sight velocity $v_{\rm{}los}$ along the minor axis of the outer ring are related by $v_{\rm{}in}=v_{\rm{}los}/\sin i$, where $i$ is the inclination of the outer ring.
For $v_{\rm{}los}\sim $2$-$4~km/s at $\sim$0.1~arcsec from the central star, the infall velocity is inferred to be 2.5$-$5~km/s at the deprojected distance of $\sim 30$~au from the central star.

From the infall velocity, we can estimate the mass infall rate $\dot{M}= 2\pi r \Sigma_{\rm{g}} v_{\rm{}in}$, where $\Sigma_{\rm{}g}$ is the gas surface density at radius $r$.
For the estimation of $\Sigma_{\rm g}$, we can employ $\rm{}^{12}CO$(2$-$1) emission.  
Because $\rm{}^{12}CO$(2$-$1) emission is most likely optically thick, this estimate provides the lower limit of the gas surface density, and thus the lower limit of the mass infall rate.  
\Figref{sigma12} shows the radial distribution of gas surface density derived from $\rm{}^{12}CO$(2$-$1) assuming the temperature expressed by equation~(5) and the gas surface density at $\sim 0.1$ arcsec from the central star is $\sim 0.003$~g/cm$^{-2}$ for the standard CO/H$_2$ ratio of $10^{-4}$.  
This is significantly lower than the gas surface density using $^{13}$CO data (\figref{sigma_g}), and the actual gas surface density can even be higher if CO is depleted (see section~3.3.2).  
However, this gas surface density results in an estimate of the mass infall rate of $\sim 10^{-8}~M_{\odot}$/yr at $r\sim 30$~au from the central star, which is considerably higher than the stellar mass accretion rate of $\sim 3.89 \times 10^{-10}~M_{\odot}$/yr derived from VLT/X-shooter spectra (\cite{manara2017}).

The evolution of protoplanetary disks is significantly affected by magnetic flux distribution (\cite{tsukamoto2015a}; \cite{tsukamoto2015b}; \cite{Bai2017}), and the wind-driven accretion by magnetohydrodynamic (MHD) effects play an important role in the mass loss process and angular momentum transport (\cite{Bai2013}; \cite{suzuki2016}).
One possible scenario that explains the velocity distortion caused by radial flow while maintaining a low stellar mass accretion rate is the combination of fast accretion only in the upper layer of the disk and disk wind.  
The mass accretion rate is low in the close vicinity of the star, probably because the accreting materials in the upper layer of the disk are loaded on magnetic field lines, and are disposed as disk wind (e.g., \cite{Sheikhnezami2012}). 
However, the wind mass loss rate is probably a few times the stellar mass accretion rate at most (\cite{hasegawa2021}), and therefore, it may be difficult to load almost all the accreting material from the outer disk onto the disk wind.

\subsection{Warp}

If the system has a warp structure, we expected the existence of a misaligned inner dust disk (\cite{casassus2018}).
In this section, we assumed that the disk system has a misalignment between the inner and outer rings.
The PA of the disk major axis should be along the direction with the highest line-of-site velocity gradient; therefore, we assumed that the major axis of the inner ring is along the PA of $-$40$^\circ$ (shown in \figref{mom1_zoom}a).

\Figref{dust_model}a shows the residual intensity map shown in \figref{visres}a projected onto the plane of the sky using the outer ring inclination (51.1$^\circ$) and position angle ($-$15.4$^\circ$).
As shown in the figure, the residual pattern is brightest in the northeast direction, which is similar to the direction along the minor axis of the inner disk, as suggested by the velocity gradient.
A misaligned inner disk can cast a shadow onto the outer disk, causing azimuthally asymmetric illumination (\cite{marino2015}).
This asymmetric illumination may cause temperature variations on the ring, thereby causing asymmetry of the dust continuum emission.  
Our observations show that the degree of azimuthal asymmetry along the outer ring is only $\sim15\%$.  
This suggests that the difference in temperature in the bright and dark regions is only a few Kelvin, while the average temperature of the outer ring is $\sim 30$~K.  
This temperature difference may be caused by the asymmetric illumination pattern caused by the inner disk.  

Here, we investigate the shadow cast by the inner disk using a simplified model made with the 3D graphics software Blender. 
We considered two optically thick misaligned rings sharing the same center where a point light source is placed (\figref{dust_model}b).
The outer ring has the same position angle and inclination as observed (PA =$-$15.4$^\circ$, inc.=51.1$^\circ$), while the inner ring has a position angle of $-$40$^\circ$, which is indicated by the observed velocity distortion.  
Thereafter, we varied the inclination of the inner ring to investigate the changes in the illumination pattern on the outer disk.  
We set the radius ratio between the outer and inner torus to 102:3 and the scale height of the two rings relative to their radius to be 0.2.
We found an illumination pattern similar to \figref{dust_model}a when the inner disk has an inclination of 70–90$^\circ$(\figref{dust_model}b). 
The northeast part of the outer ring is illuminated, and the temperature of this part may be higher than that of the other parts.  
Moreover, the inner disk with an inclination of 70$–$90$^\circ$ is consistent with the observed velocity distortion of a few km/s, because the Keplerian velocity at $r \sim 0.1$~arcsec $\approx20$~au along the major axis of the inner disk is 6~km/s.

Therefore, the misaligned inner disk is a plausible interpretation of the velocity distribution and dust asymmetry.  
To obtain a realistic temperature distribution, rather than an illumination pattern, we need full radiative transfer modeling, which will be presented in a future work. 

Detailed study of the warped inner disk is left as a future work.
As one of the approach, we propose conducting an observation of ro-vibrational CO line emission at 4.7 $\mu$m for SY Cha. 
This line is formed in the heated surfaces of the disk at radii of 0.1$-$10 au (\cite{brown2013}), and the compact inner gas disk thus is thus expected to be traced.

\citet{pontoppidan2008} observed a CO 4.7 $\mu$m line in protoplanetary disks around SR 21, HD 135344B and TW Hya using the CRIRES high-resolution infrared spectrometer on the Very Large Telescope (VLT) and estimated the PA and inclination of the inner disks to within a few degrees, though with a caveat that these estimates listed by \citet{pontoppidan2008} rely on accurate estimation of the stellar mass, which was not well constrained at the time.

\section{Conclusions} 

We reported ALMA observations of the transitional disk around SY Cha.
The dust continuum emission and the $J=$2$-$1 line emission of CO isotopologues at Band 6 were observed.
Our conclusions are summarized as follows;

\begin{enumerate}
\item The 225 GHz dust continuum emission shows a clear ring structure with a radius of $\sim$100 au and central point sources.
The total dust mass is estimated to be $1.0\times10^{-4}M_{\odot}$.
We suggested that the central point source is actually a ring because the SED does not show a significant IR excess.
From visibility-domain analyses, we found extended emission outside the outer ring and weak asymmetry in the outer ring.  
The asymmetry shows that the ring is bright in the north-east direction.

\item We detected $^{12}$CO, $^{13}$CO and C$^{18}$O emission.  
The $^{13}$CO emission is concentrated in the vicinity of the dust ring, while $^{12}$CO emission was detected inside the dust ring, indicating that there is some gas remaining inside the dust ring.  
Assuming that $^{12}$CO is a tracer of disk temperature, we estimated the total gas mass as $2.2\times10^{-4}M_{\odot}$ and the peak column density of the disk gas as $\sim 0.025$~g/cm$^2$ if CO is not depleted.
The estimated gas value may be increased by a factor of 10$-$100 by CO depletion.

\item The moment 1 map of the CO line emission shows that the velocity pattern agree with the Keplerian rotation around 0.78 $M_{\odot}$; however, the velocity pattern in the vicinity of the central star is distorted.
The origin of the velocity distortion is yet inconclusive. 
If the velocity distortion is caused by radial flow, the mass inflow rate at $\sim 30$~au is well above the stellar mass accretion rate, indicating that a significant amount of mass is being ejected as disk wind.  
If the velocity distortion is caused by a warped inner disk, we expected that the inner disk is close to edge-on (inclination of 70–90$^\circ$).  
The weak asymmetry of the outer disk is consistent with the asymmetric illumination of the central star caused by the warped inner disk.
\end{enumerate}

\begin{ack}
This study makes use of the following ALMA data: ADS/JAO.ALMA\#2018.1.00689.S.
ALMA is a partnership of ESO (representing its member states), NSF (USA) and NINS (Japan), together with NRC (Canada), MOST and ASIAA (Taiwan), and KASI (Republic of Korea), in cooperation with the Republic of Chile. 
The Joint ALMA Observatory is operated by ESO, AUI/NRAO and NAOJ. 
The National Radio Astronomy Observatory is a facility of the National Science Foundation, operated under cooperative agreement by Associated Universities, Inc.
This work was supported by JSPS KAKENHI grant numbers 17H01103 and 18H05441. This paper makes use of the following ALMA data: ADS/JAO.ALMA\#2018.1.00689.S. ALMA is a partnership of ESO (representing its member states), NSF (USA) and NINS (Japan), together with NRC (Canada), MOST and ASIAA (Taiwan), and KASI (Republic of Korea), in cooperation with the Republic of Chile. The Joint ALMA Observatory is operated by ESO, AUI/NRAO and NAOJ.
Y.Y. was supported by the NAOJ ALMA Scientific Research grant number 2019-12A.
Y.H. was supported by the Jet Propulsion Laboratory, California Institute of Technology, under a contract with the National Aeronautics and Space Administration (80NM0018D0004).
T.T. was supported by JSPS KAKENHI grant number 20K04017.
\end{ack}

\textit{Software: astropy, CASA, Numpy, Matplotlib, emcee, keplerian\_mask, bettermoments, Blender}
\appendix

\section{Channel maps}
\begin{figure*}[h]
   \begin{center}
     \includegraphics[width=\textwidth]{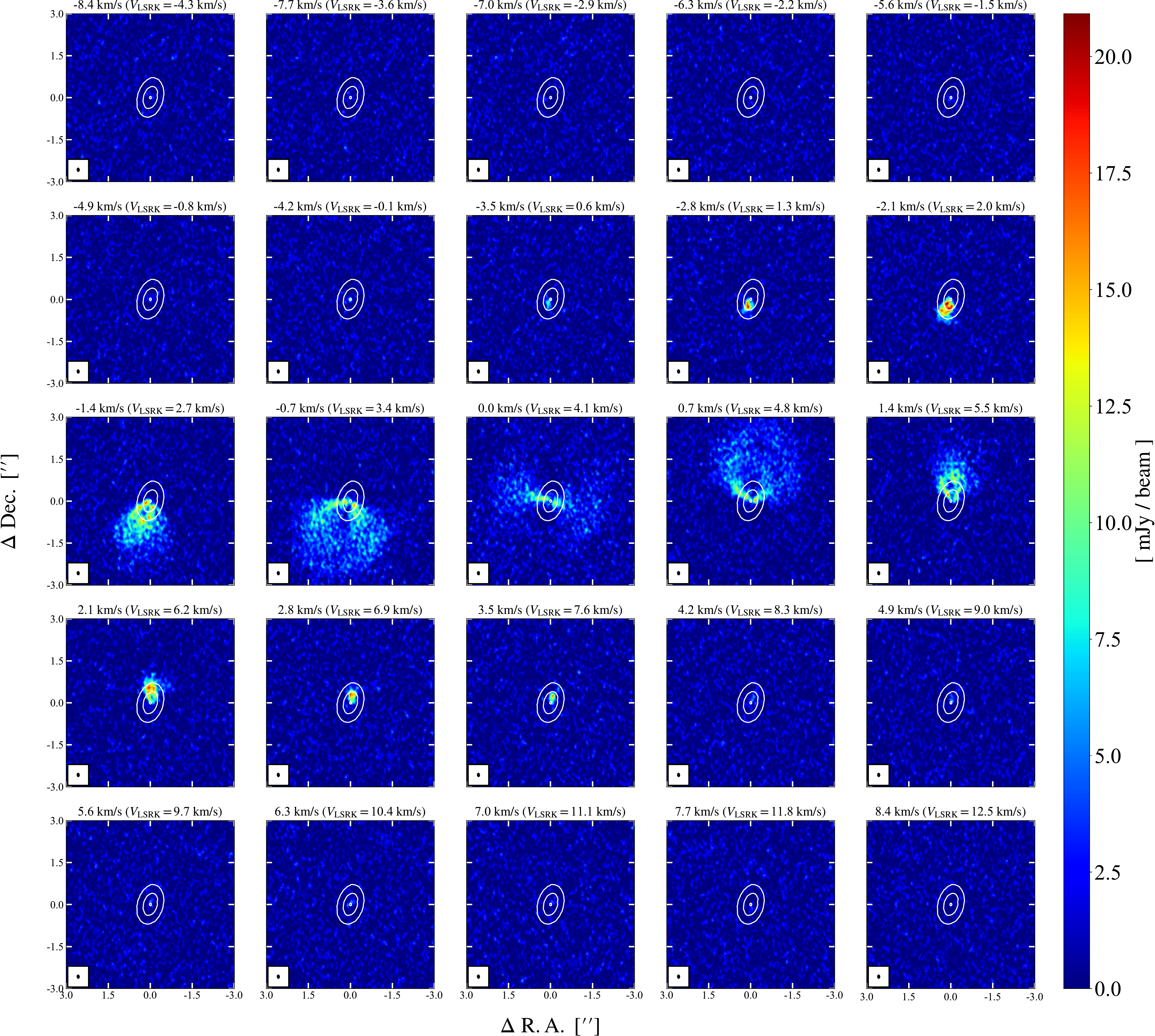}%
   \end{center}
   \caption{$\rm{}^{12}CO$(2$-$1) Channel Map in the concatenated data.
   The black ellipse at the bottom left of each panel represents the beam size.
   The contours in these maps indicate the 20$\sigma$ level of the dust continuum, where $1\sigma = 14.9\ \rm{}\mu Jy/beam$.
   }
\label{fig:chan_12CO}
\end{figure*}

\begin{figure*}[h]
   \begin{center}
     \includegraphics[width=\textwidth]{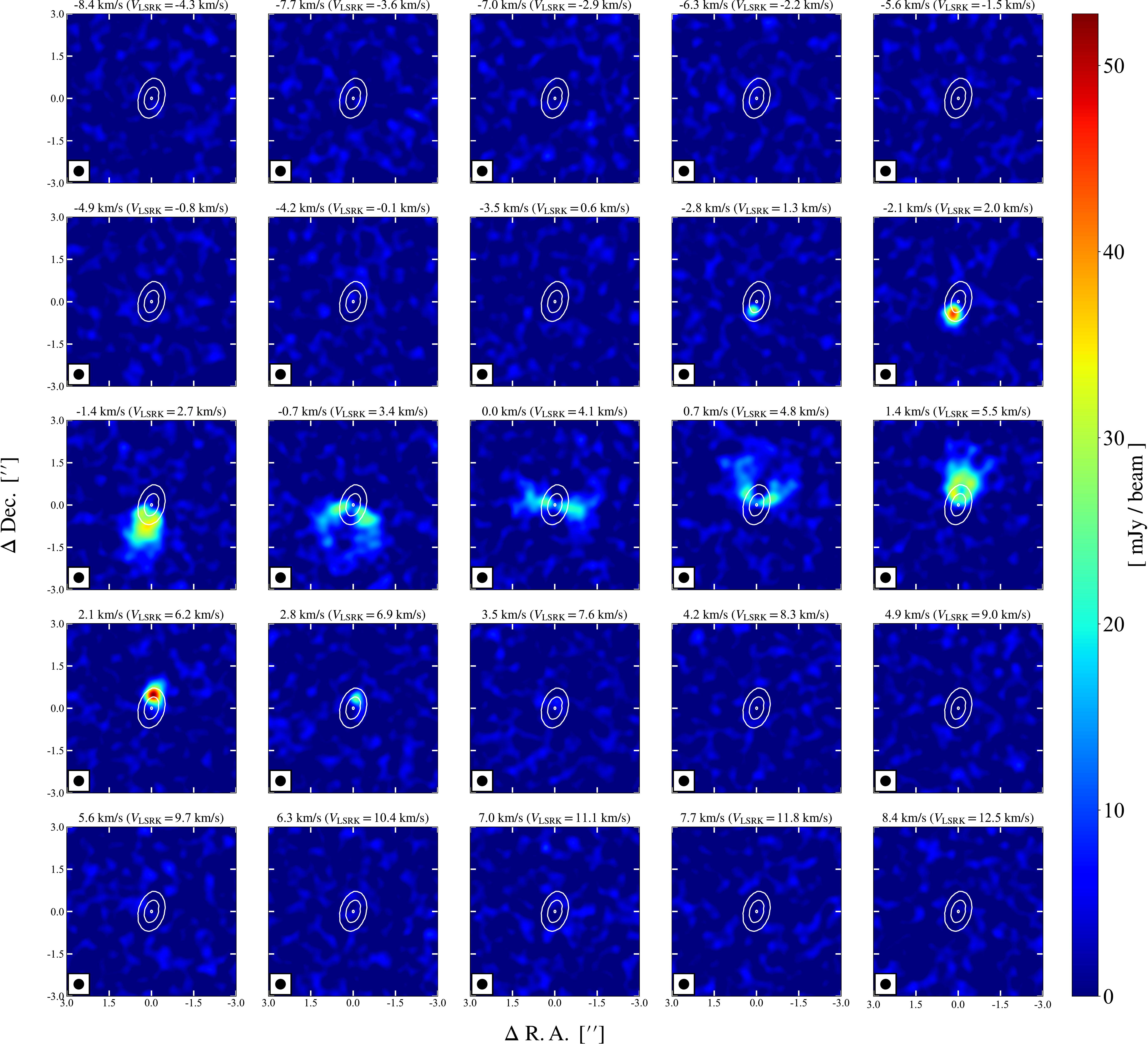}%
   \end{center}
   \caption{$\rm{}^{13}CO$(2$-$1) Channel Map in the CC data.
   The black ellipse at the bottom left of each panel represents the beam size.
   The contours in these maps indicate the 20$\sigma$ level of the dust continuum, where $1\sigma = 14.9\ \rm{}\mu Jy/beam$.
   }
\label{fig:chan_13CO}
\end{figure*}

\begin{figure*}[h]
   \begin{center}
     \includegraphics[width=\textwidth]{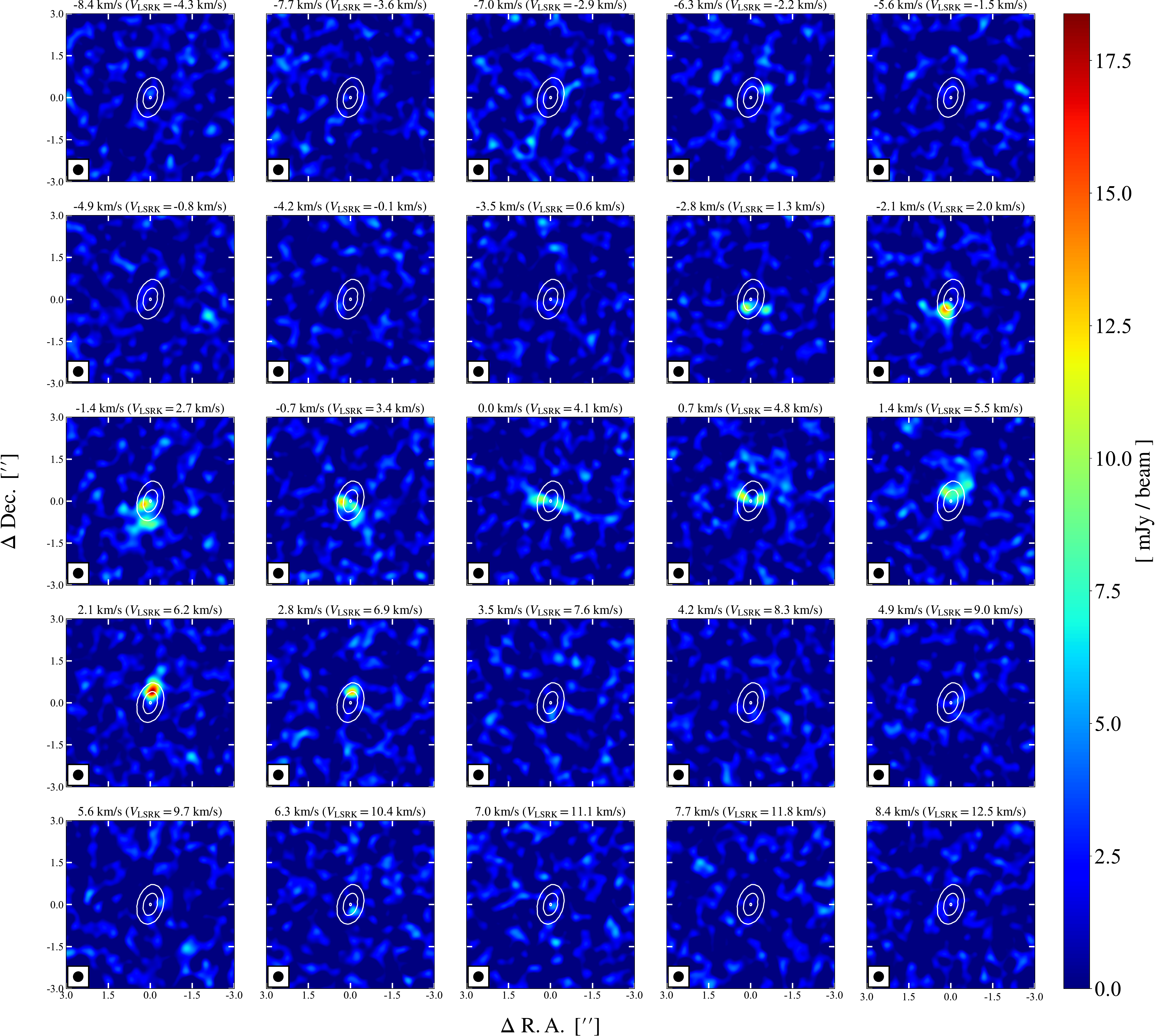}%
   \end{center}
   \caption{$\rm{}C^{18}O$(2$-$1) Channel Map in the CC data.
   The black ellipse at the bottom left of each panel represents the beam size.
   The contours in these maps indicate the 20$\sigma$ level of the dust continuum, where $1\sigma = 14.9\ \rm{}\mu Jy/beam$.
   }
\label{fig:chan_C18O}
\end{figure*}

Figures \figr{chan_12CO}$-$\figr{chan_C18O} show the channel maps of the CO line emission at $V_{\rm{LSRK}}=-4.3$ to $+12.5$ km/s.
See \tabref{imageparam} for information of these channel maps.

\section{Calculation of molecular column density}

Calculating the optical depth $\tau_{\rm{g}}$ using equation~(8), the molecular spectral line column density is expressed by the following equations (\cite{mangum2015}):

\begin{eqnarray}
    N_{\rm{}tot}=&\frac{3h}{8\pi^3|\mu_{\rm{}lu}|^2}\frac{Q_{\rm{}tot}}{g_{\rm{}u}}\exp\Big(\frac{E_{\rm{}u}}{kT_{\rm{}ex}}\Big) \nonumber \\
    &\times\Big[\exp\Big(\frac{h\nu}{kT_{\rm{}ex}}\Big)-1\Big]^{-1}\int\tau_{\rm{}g}dv,
\end{eqnarray}
\begin{eqnarray}
    g_{\rm{}u}=2J_{\rm{}u}+1,
\end{eqnarray}
\begin{eqnarray}
    E_{\rm{}u}=hBJ_{\rm{}u}(J_{\rm{}u}+1),
\end{eqnarray}
\begin{eqnarray}
  Q_{\rm{}tot}\approx \frac{kT_{\rm{g}}}{hB}+\frac{1}{3},
\end{eqnarray}
\begin{eqnarray}
    |\mu_{\rm{}lu}|^2=\mu^2S,
\end{eqnarray}
\begin{eqnarray}
    S=\frac{J_{\rm{}u}^2}{J_{\rm{}u}(2J_{\rm{}u}+1)},
\end{eqnarray}
where $\mu$, $J_{\rm{u}}$, and $B$ are the dipole moment, rotational quantum number of the line emission upper level, and rigid rotor rotational constant, respectively (parameters of each CO isotopologue are shown in \tabref{colist}).
The excitation temperature $T_{\rm{ex}}$ is equal to $T_{\rm{g}}$ in the LTE analysis.
The gas column density is then calculated by
\begin{eqnarray}
    \Sigma_{\rm{}g} =\frac{m_{\rm{H_2}}N_{\rm{tot}}}{X},
\end{eqnarray}
where $m_{\rm{H_2}}$($=3.32\times10^{-24}\rm{g}$) is the molecular mass of $\rm{H_2}$, and $X$ is the molecular abundance ratio with $\rm{}H_2$.

In this study, the following equation was used to calculate the total flux of the $\rm{}C^{18}O$(2$-$1) line emission:

\begin{eqnarray}
N_{\rm{}tot}({\rm{C^{18}O}})=&\frac{3c^2}{16\pi^3 \Omega_S S\mu^2\nu^3}\Big(\frac{Q_{\rm{rot}}}{g_{\rm{u}}}\Big) \nonumber \\
&\times \exp{\Big(\frac{E_{\rm{u}}}{kT}\Big)}\int S_{\nu}dv ,
\end{eqnarray}
where $S_\nu$ and $\Omega_S$ are the flux density and beam size, respectively.

\begin{table}[t]
\tbl{Physical parameters of CO isotopologue}{%
\begin{tabular}{l|ccc} 
    \hline
                     & $\rm{}^{12}CO$(2$-$1) & $\rm{}^{13}CO$(2$-$1) & $\rm{}C^{18}O$(2$-$1)\\
    \hline
    $\mu$ [$10^{-18}$ esu cm]& 0.110& 0.110 & 0.1098 \\
    $J_{\rm{}u}$& 2 & 2 &2\\
    $B$ [MHz] & 57635.96& 55101.01 & 54891.42\\
    \shortstack[l]{abundance ratio \\ \hspace{30pt} with $\rm{}H_2$ }& $10^{-4}$ &  $1/67\times10^{-4}$ & $1/444\times10^{-4}$\\
    \hline
\end{tabular}}\label{tab:colist}
\end{table}

\section{Corner plots of MCMC fitting results}

\begin{figure*}[h]
   \begin{center}
     \includegraphics[width=\textwidth]{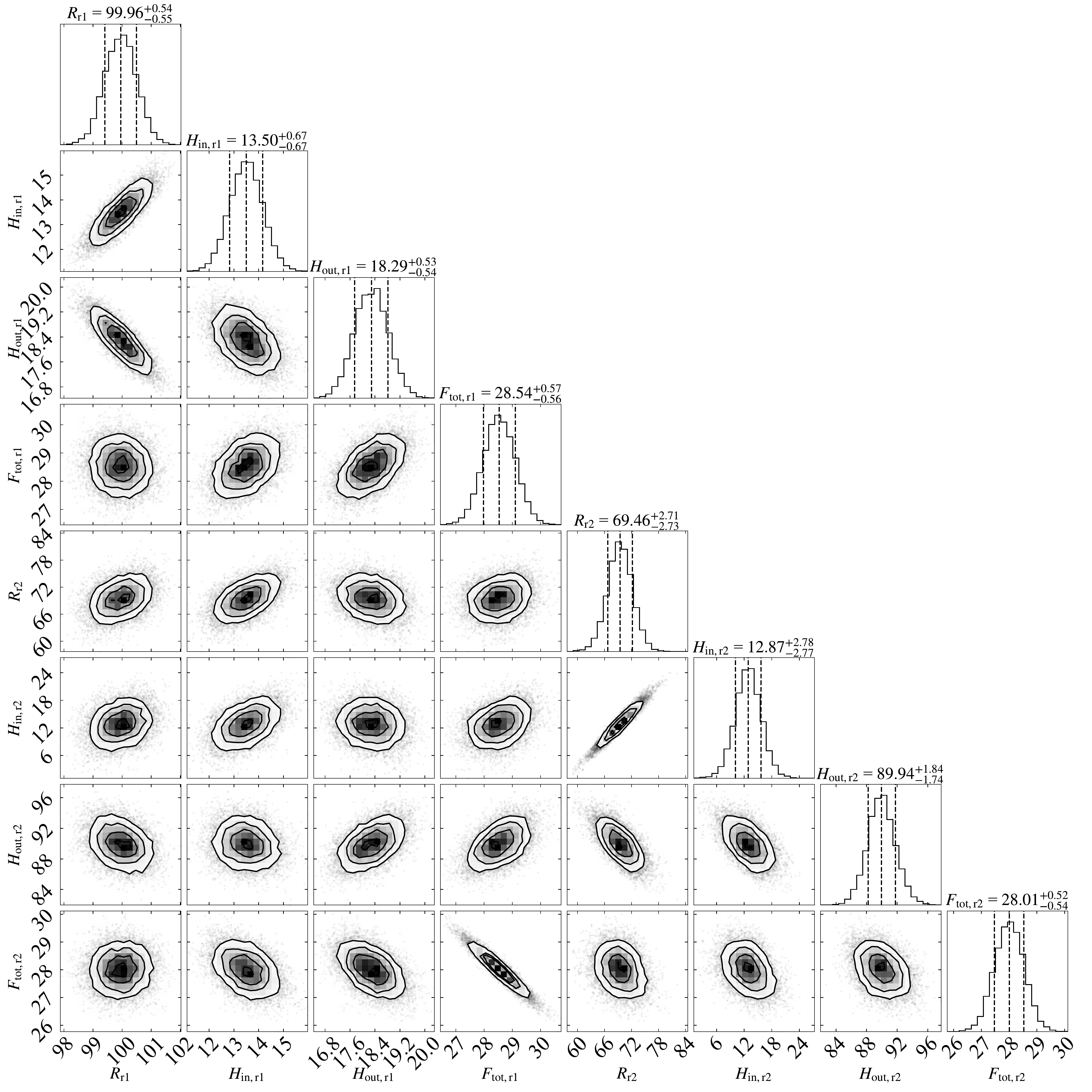}%
   \end{center}
   \caption{Corner plot of MCMC results in visibility analysis.
   The median values and 1$\sigma$ of the best-fit parameters are shown as vertical dashed lines.
   }
   \label{fig:corner_two}
\end{figure*}

\Figref{corner_two} shows corner plot of the MCMC posteriors calculated in visibility fitting given the model presented as equation~(2).
The plot shows the posterior sampling provided by the last 4000 steps of the 200 walkers chain. 

\clearpage
\bibliography{thesis}{} 

\begin{thebibliography}{}
\expandafter\ifx\csname natexlab\endcsname\relax\def\natexlab#1{#1}\fi

\bibitem[{{ALMA Partnership} {et~al.}(2015){ALMA Partnership}, {Brogan},
  {P{\'e}rez}, {Hunter}, {Dent}, {Hales}, {Hills}, {Corder}, {Fomalont},
  {Vlahakis}, {Asaki}, {Barkats}, {Hirota}, {Hodge}, {Impellizzeri}, {Kneissl},
  {Liuzzo}, {Lucas}, {Marcelino}, {Matsushita}, {Nakanishi}, {Phillips},
  {Richards}, {Toledo}, {Aladro}, {Broguiere}, {Cortes}, {Cortes}, {Espada},
  {Galarza}, {Garcia-Appadoo}, {Guzman-Ramirez}, {Humphreys}, {Jung}, {Kameno},
  {Laing}, {Leon}, {Marconi}, {Mignano}, {Nikolic}, {Nyman}, {Radiszcz},
  {Remijan}, {Rod{\'o}n}, {Sawada}, {Takahashi}, {Tilanus}, {Vila Vilaro},
  {Watson}, {Wiklind}, {Akiyama}, {Chapillon}, {de Gregorio-Monsalvo}, {Di
  Francesco}, {Gueth}, {Kawamura}, {Lee}, {Nguyen Luong}, {Mangum}, {Pietu},
  {Sanhueza}, {Saigo}, {Takakuwa}, {Ubach}, {van Kempen}, {Wootten},
  {Castro-Carrizo}, {Francke}, {Gallardo}, {Garcia}, {Gonzalez}, {Hill},
  {Kaminski}, {Kurono}, {Liu}, {Lopez}, {Morales}, {Plarre}, {Schieven},
  {Testi}, {Videla}, {Villard}, {Andreani}, {Hibbard}, \&
  {Tatematsu}}]{alma2015}
{ALMA Partnership}, {et~al.} 2015, \apjl, 808, L3

\bibitem[{{Andrews} {et~al.}(2016){Andrews}, {Wilner}, {Zhu}, {Birnstiel},
  {Carpenter}, {P{\'e}rez}, {Bai}, {{\"O}berg}, {Hughes}, {Isella}, \&
  {Ricci}}]{andrews2016}
{Andrews}, S.~M., {et~al.} 2016, \apjl, 820, L40

\bibitem[{Baddour(2009)}]{Baddour_2009}
Baddour, N. 2009, J. Opt. Soc. Am. A, 26, 1767

\bibitem[{{Bai} \& {Stone}(2013)}]{Bai2013}
{Bai}, X.-N., \& {Stone}, J.~M. 2013, \apj, 769, 76

\bibitem[{{Bai} \& {Stone}(2017)}]{Bai2017}
---. 2017, \apj, 836, 46

\bibitem[{{Bayo} {et~al.}(2008){Bayo}, {Rodrigo}, {Barrado Y Navascu{\'e}s},
  {Solano}, {Guti{\'e}rrez}, {Morales-Calder{\'o}n}, \& {Allard}}]{vosa2008}
{Bayo}, A., {Rodrigo}, C., {Barrado Y Navascu{\'e}s}, D., {Solano}, E.,
  {Guti{\'e}rrez}, R., {Morales-Calder{\'o}n}, M., \& {Allard}, F. 2008, \aap,
  492, 277

\bibitem[{{Beckwith} {et~al.}(1990){Beckwith}, {Sargent}, {Chini}, \&
  {Guesten}}]{beckwith1990}
{Beckwith}, S. V.~W., {Sargent}, A.~I., {Chini}, R.~S., \& {Guesten}, R. 1990,
  \aj, 99, 924

\bibitem[{{Belloche} {et~al.}(2011){Belloche}, {Schuller}, {Parise},
  {Andr{\'e}}, {Hatchell}, {J{\o}rgensen}, {Bontemps}, {Wei{\ss}}, {Menten}, \&
  {Muders}}]{APEX2011}
{Belloche}, A., {et~al.} 2011, \aap, 527, A145

\bibitem[{{Brown} {et~al.}(2013){Brown}, {Pontoppidan}, {van Dishoeck},
  {Herczeg}, {Blake}, \& {Smette}}]{brown2013}
{Brown}, J.~M., {Pontoppidan}, K.~M., {van Dishoeck}, E.~F., {Herczeg}, G.~J.,
  {Blake}, G.~A., \& {Smette}, A. 2013, \apj, 770, 94

\bibitem[{{Calvet} {et~al.}(2002){Calvet}, {D'Alessio}, {Hartmann}, {Wilner},
  {Walsh}, \& {Sitko}}]{Calvet2002}
{Calvet}, N., {D'Alessio}, P., {Hartmann}, L., {Wilner}, D., {Walsh}, A., \&
  {Sitko}, M. 2002, \apj, 568, 1008

\bibitem[{{Calvet} {et~al.}(2005){Calvet}, {D'Alessio}, {Watson},
  {Franco-Hern{\'a}ndez}, {Furlan}, {Green}, {Sutter}, {Forrest}, {Hartmann},
  {Uchida}, {Keller}, {Sargent}, {Najita}, {Herter}, {Barry}, \&
  {Hall}}]{Calvet2005}
{Calvet}, N., {et~al.} 2005, \apjl, 630, L185

\bibitem[{{Casassus} {et~al.}(2015){Casassus}, {Marino}, {P{\'e}rez}, {Roman},
  {Dunhill}, {Armitage}, {Cuadra}, {Wootten}, {van der Plas}, {Cieza}, {Moral},
  {Christiaens}, \& {Montesinos}}]{casassus2015}
{Casassus}, S., {et~al.} 2015, \apj, 811, 92

\bibitem[{{Casassus} {et~al.}(2018){Casassus}, {Avenhaus}, {P{\'e}rez},
  {Navarro}, {C{\'a}rcamo}, {Marino}, {Cieza}, {Quanz}, {Alarc{\'o}n}, {Zurlo},
  {Osses}, {Rannou}, {Rom{\'a}n}, \& {Barraza}}]{casassus2018}
---. 2018, \mnras, 477, 5104

\bibitem[{{Chiang} \& {Goldreich}(1997)}]{Chiang1997}
{Chiang}, E.~I., \& {Goldreich}, P. 1997, \apj, 490, 368

\bibitem[{{Cornwell}(2008)}]{cornwell2008}
{Cornwell}, T.~J. 2008, IEEE Journal of Selected Topics in Signal Processing,
  2, 793

\bibitem[{{Cutri} {et~al.}(2003){Cutri}, {Skrutskie}, {van Dyk}, {Beichman},
  {Carpenter}, {Chester}, {Cambresy}, {Evans}, {Fowler}, {Gizis}, {Howard},
  {Huchra}, {Jarrett}, {Kopan}, {Kirkpatrick}, {Light}, {Marsh}, {McCallon},
  {Schneider}, {Stiening}, {Sykes}, {Weinberg}, {Wheaton}, {Wheelock}, \&
  {Zacarias}}]{2MASS2003}
{Cutri}, R.~M., {et~al.} 2003, VizieR Online Data Catalog, II/246

\bibitem[{{de Gregorio-Monsalvo} {et~al.}(2013){de Gregorio-Monsalvo},
  {M{\'e}nard}, {Dent}, {Pinte}, {L{\'o}pez}, {Klaassen}, {Hales},
  {Cort{\'e}s}, {Rawlings}, {Tachihara}, {Testi}, {Takahashi}, {Chapillon},
  {Mathews}, {Juhasz}, {Akiyama}, {Higuchi}, {Saito}, {Nyman}, {Phillips},
  {Rod{\'o}n}, {Corder}, \& {Van Kempen}}]{gregorio2013}
{de Gregorio-Monsalvo}, I., {et~al.} 2013, \aap, 557, A133

\bibitem[{{Dong} {et~al.}(2015){Dong}, {Zhu}, \& {Whitney}}]{dong15gap}
{Dong}, R., {Zhu}, Z., \& {Whitney}, B. 2015, \apj, 809, 93

\bibitem[{{Espaillat} {et~al.}(2011){Espaillat}, {Furlan}, {D'Alessio},
  {Sargent}, {Nagel}, {Calvet}, {Watson}, \& {Muzerolle}}]{espaillat2011}
{Espaillat}, C., {Furlan}, E., {D'Alessio}, P., {Sargent}, B., {Nagel}, E.,
  {Calvet}, N., {Watson}, D.~M., \& {Muzerolle}, J. 2011, \apj, 728, 49

\bibitem[{{Foreman-Mackey} {et~al.}(2013){Foreman-Mackey}, {Hogg}, {Lang}, \&
  {Goodman}}]{foreman2013}
{Foreman-Mackey}, D., {Hogg}, D.~W., {Lang}, D., \& {Goodman}, J. 2013, \pasp,
  125, 306

\bibitem[{{Fouqu{\'e}} {et~al.}(2000){Fouqu{\'e}}, {Chevallier}, {Cohen},
  {Galliano}, {Loup}, {Alard}, {de Batz}, {Bertin}, {Borsenberger}, {Cioni},
  {Copet}, {Dennefeld}, {Derriere}, {Deul}, {Duc}, {Egret}, {Epchtein},
  {Forveille}, {Garz{\'o}n}, {Habing}, {Hron}, {Kimeswenger}, {Lacombe}, {Le
  Bertre}, {Mamon}, {Omont}, {Paturel}, {Pau}, {Persi}, {Robin}, {Rouan},
  {Schultheis}, {Simon}, {Tiph{\`e}ne}, {Vauglin}, \& {Wagner}}]{DENIS2000}
{Fouqu{\'e}}, P., {et~al.} 2000, \aaps, 141, 313

\bibitem[{{Francis} \& {van der Marel}(2020)}]{francis2020}
{Francis}, L., \& {van der Marel}, N. 2020, \apj, 892, 111

\bibitem[{{Frasca} {et~al.}(2015){Frasca}, {Biazzo}, {Lanzafame}, {Alcal{\'a}},
  {Brugaletta}, {Klutsch}, {Stelzer}, {Sacco}, {Spina}, {Jeffries}, {Montes},
  {Alfaro}, {Barentsen}, {Bonito}, {Gameiro}, {L{\'o}pez-Santiago}, {Pace},
  {Pasquini}, {Prisinzano}, {Sousa}, {Gilmore}, {Randich}, {Micela},
  {Bragaglia}, {Flaccomio}, {Bayo}, {Costado}, {Franciosini}, {Hill},
  {Hourihane}, {Jofr{\'e}}, {Lardo}, {Maiorca}, {Masseron}, {Morbidelli}, \&
  {Worley}}]{gaia2015}
{Frasca}, A., {et~al.} 2015, \aap, 575, A4

\bibitem[{{Fukagawa} {et~al.}(2013){Fukagawa}, {Tsukagoshi}, {Momose}, {Saigo},
  {Ohashi}, {Kitamura}, {Inutsuka}, {Muto}, {Nomura}, {Takeuchi}, {Kobayashi},
  {Hanawa}, {Akiyama}, {Honda}, {Fujiwara}, {Kataoka}, {Takahashi}, \&
  {Shibai}}]{fukagawa2013}
{Fukagawa}, M., {et~al.} 2013, \pasj, 65, L14

\bibitem[{{Gaia Collaboration} {et~al.}(2018){Gaia Collaboration}, {Brown},
  {Vallenari}, {Prusti}, {de Bruijne}, {Babusiaux}, {Bailer-Jones}, {Biermann},
  {Evans}, {Eyer}, {Jansen}, {Jordi}, {Klioner}, {Lammers}, {Lindegren},
  {Luri}, {Mignard}, {Panem}, {Pourbaix}, {Randich}, {Sartoretti}, {Siddiqui},
  {Soubiran}, {van Leeuwen}, {Walton}, {Arenou}, {Bastian}, {Cropper},
  {Drimmel}, {Katz}, {Lattanzi}, {Bakker}, {Cacciari}, {Casta{\~n}eda},
  {Chaoul}, {Cheek}, {De Angeli}, {Fabricius}, {Guerra}, {Holl}, {Masana},
  {Messineo}, {Mowlavi}, {Nienartowicz}, {Panuzzo}, {Portell}, {Riello},
  {Seabroke}, {Tanga}, {Th{\'e}venin}, {Gracia-Abril}, {Comoretto},
  {Garcia-Reinaldos}, {Teyssier}, {Altmann}, {Andrae}, {Audard},
  {Bellas-Velidis}, {Benson}, {Berthier}, {Blomme}, {Burgess}, {Busso},
  {Carry}, {Cellino}, {Clementini}, {Clotet}, {Creevey}, {Davidson}, {De
  Ridder}, {Delchambre}, {Dell'Oro}, {Ducourant},
  {Fern{\'a}ndez-Hern{\'a}ndez}, {Fouesneau}, {Fr{\'e}mat}, {Galluccio},
  {Garc{\'\i}a-Torres}, {Gonz{\'a}lez-N{\'u}{\~n}ez}, {Gonz{\'a}lez-Vidal},
  {Gosset}, {Guy}, {Halbwachs}, {Hambly}, {Harrison}, {Hern{\'a}ndez},
  {Hestroffer}, {Hodgkin}, {Hutton}, {Jasniewicz}, {Jean-Antoine-Piccolo},
  {Jordan}, {Korn}, {Krone-Martins}, {Lanzafame}, {Lebzelter}, {L{\"o}ffler},
  {Manteiga}, {Marrese}, {Mart{\'\i}n-Fleitas}, {Moitinho}, {Mora}, {Muinonen},
  {Osinde}, {Pancino}, {Pauwels}, {Petit}, {Recio-Blanco}, {Richards},
  {Rimoldini}, {Robin}, {Sarro}, {Siopis}, {Smith}, {Sozzetti}, {S{\"u}veges},
  {Torra}, {van Reeven}, {Abbas}, {Abreu Aramburu}, {Accart}, {Aerts},
  {Altavilla}, {{\'A}lvarez}, {Alvarez}, {Alves}, {Anderson}, {Andrei},
  {Anglada Varela}, {Antiche}, {Antoja}, {Arcay}, {Astraatmadja}, {Bach},
  {Baker}, {Balaguer-N{\'u}{\~n}ez}, {Balm}, {Barache}, {Barata}, {Barbato},
  {Barblan}, {Barklem}, {Barrado}, {Barros}, {Barstow}, {Bartholom{\'e}
  Mu{\~n}oz}, {Bassilana}, {Becciani}, {Bellazzini}, {Berihuete}, {Bertone},
  {Bianchi}, {Bienaym{\'e}}, {Blanco-Cuaresma}, {Boch}, {Boeche}, {Bombrun},
  {Borrachero}, {Bossini}, {Bouquillon}, {Bourda}, {Bragaglia}, {Bramante},
  {Breddels}, {Bressan}, {Brouillet}, {Br{\"u}semeister}, {Brugaletta},
  {Bucciarelli}, {Burlacu}, {Busonero}, {Butkevich}, {Buzzi}, {Caffau},
  {Cancelliere}, {Cannizzaro}, {Cantat-Gaudin}, {Carballo}, {Carlucci},
  {Carrasco}, {Casamiquela}, {Castellani}, {Castro-Ginard}, {Charlot},
  {Chemin}, {Chiavassa}, {Cocozza}, {Costigan}, {Cowell}, {Crifo}, {Crosta},
  {Crowley}, {Cuypers}, {Dafonte}, {Damerdji}, {Dapergolas}, {David}, {David},
  {de Laverny}, {De Luise}, {De March}, {de Martino}, {de Souza}, {de Torres},
  {Debosscher}, {del Pozo}, {Delbo}, {Delgado}, {Delgado}, {Di Matteo},
  {Diakite}, {Diener}, {Distefano}, {Dolding}, {Drazinos}, {Dur{\'a}n},
  {Edvardsson}, {Enke}, {Eriksson}, {Esquej}, {Eynard Bontemps}, {Fabre},
  {Fabrizio}, {Faigler}, {Falc{\~a}o}, {Farr{\`a}s Casas}, {Federici},
  {Fedorets}, {Fernique}, {Figueras}, {Filippi}, {Findeisen}, {Fonti},
  {Fraile}, {Fraser}, {Fr{\'e}zouls}, {Gai}, {Galleti}, {Garabato},
  {Garc{\'\i}a-Sedano}, {Garofalo}, {Garralda}, {Gavel}, {Gavras}, {Gerssen},
  {Geyer}, {Giacobbe}, {Gilmore}, {Girona}, {Giuffrida}, {Glass}, {Gomes},
  {Granvik}, {Gueguen}, {Guerrier}, {Guiraud}, {Guti{\'e}rrez-S{\'a}nchez},
  {Haigron}, {Hatzidimitriou}, {Hauser}, {Haywood}, {Heiter}, {Helmi}, {Heu},
  {Hilger}, {Hobbs}, {Hofmann}, {Holland}, {Huckle}, {Hypki}, {Icardi},
  {Jan{\ss}en}, {Jevardat de Fombelle}, {Jonker}, {Juh{\'a}sz}, {Julbe},
  {Karampelas}, {Kewley}, {Klar}, {Kochoska}, {Kohley}, {Kolenberg},
  {Kontizas}, {Kontizas}, {Koposov}, {Kordopatis}, {Kostrzewa-Rutkowska},
  {Koubsky}, {Lambert}, {Lanza}, {Lasne}, {Lavigne}, {Le Fustec}, {Le
  Poncin-Lafitte}, {Lebreton}, {Leccia}, {Leclerc}, {Lecoeur-Taibi},
  {Lenhardt}, {Leroux}, {Liao}, {Licata}, {Lindstr{\o}m}, {Lister}, {Livanou},
  {Lobel}, {L{\'o}pez}, {Managau}, {Mann}, {Mantelet}, {Marchal}, {Marchant},
  {Marconi}, {Marinoni}, {Marschalk{\'o}}, {Marshall}, {Martino}, {Marton},
  {Mary}, {Massari}, {Matijevi{\v{c}}}, {Mazeh}, {McMillan}, {Messina},
  {Michalik}, {Millar}, {Molina}, {Molinaro}, {Moln{\'a}r}, {Montegriffo},
  {Mor}, {Morbidelli}, {Morel}, {Morris}, {Mulone}, {Muraveva}, {Musella},
  {Nelemans}, {Nicastro}, {Noval}, {O'Mullane}, {Ord{\'e}novic},
  {Ord{\'o}{\~n}ez-Blanco}, {Osborne}, {Pagani}, {Pagano}, {Pailler},
  {Palacin}, {Palaversa}, {Panahi}, {Pawlak}, {Piersimoni}, {Pineau}, {Plachy},
  {Plum}, {Poggio}, {Poujoulet}, {Pr{\v{s}}a}, {Pulone}, {Racero}, {Ragaini},
  {Rambaux}, {Ramos-Lerate}, {Regibo}, {Reyl{\'e}}, {Riclet}, {Ripepi}, {Riva},
  {Rivard}, {Rixon}, {Roegiers}, {Roelens}, {Romero-G{\'o}mez}, {Rowell},
  {Royer}, {Ruiz-Dern}, {Sadowski}, {Sagrist{\`a} Sell{\'e}s}, {Sahlmann},
  {Salgado}, {Salguero}, {Sanna}, {Santana-Ros}, {Sarasso}, {Savietto},
  {Schultheis}, {Sciacca}, {Segol}, {Segovia}, {S{\'e}gransan}, {Shih},
  {Siltala}, {Silva}, {Smart}, {Smith}, {Solano}, {Solitro}, {Sordo}, {Soria
  Nieto}, {Souchay}, {Spagna}, {Spoto}, {Stampa}, {Steele},
  {Steidelm{\"u}ller}, {Stephenson}, {Stoev}, {Suess}, {Surdej}, {Szabados},
  {Szegedi-Elek}, {Tapiador}, {Taris}, {Tauran}, {Taylor}, {Teixeira},
  {Terrett}, {Teyssandier}, {Thuillot}, {Titarenko}, {Torra Clotet}, {Turon},
  {Ulla}, {Utrilla}, {Uzzi}, {Vaillant}, {Valentini}, {Valette}, {van Elteren},
  {Van Hemelryck}, {van Leeuwen}, {Vaschetto}, {Vecchiato}, {Veljanoski},
  {Viala}, {Vicente}, {Vogt}, {von Essen}, {Voss}, {Votruba}, {Voutsinas},
  {Walmsley}, {Weiler}, {Wertz}, {Wevers}, {Wyrzykowski}, {Yoldas},
  {{\v{Z}}erjal}, {Ziaeepour}, {Zorec}, {Zschocke}, {Zucker}, {Zurbach}, \&
  {Zwitter}}]{gaia2018}
{Gaia Collaboration}, {et~al.} 2018, \aap, 616, A1

\bibitem[{{Gaia Collaboration} {et~al.}(2021){Gaia Collaboration}, {Brown},
  {Vallenari}, {Prusti}, {de Bruijne}, {Babusiaux}, {Biermann}, {Creevey},
  {Evans}, {Eyer}, {Hutton}, {Jansen}, {Jordi}, {Klioner}, {Lammers},
  {Lindegren}, {Luri}, {Mignard}, {Panem}, {Pourbaix}, {Randich}, {Sartoretti},
  {Soubiran}, {Walton}, {Arenou}, {Bailer-Jones}, {Bastian}, {Cropper},
  {Drimmel}, {Katz}, {Lattanzi}, {van Leeuwen}, {Bakker}, {Cacciari},
  {Casta{\~n}eda}, {De Angeli}, {Ducourant}, {Fabricius}, {Fouesneau},
  {Fr{\'e}mat}, {Guerra}, {Guerrier}, {Guiraud}, {Jean-Antoine Piccolo},
  {Masana}, {Messineo}, {Mowlavi}, {Nicolas}, {Nienartowicz}, {Pailler},
  {Panuzzo}, {Riclet}, {Roux}, {Seabroke}, {Sordo}, {Tanga}, {Th{\'e}venin},
  {Gracia-Abril}, {Portell}, {Teyssier}, {Altmann}, {Andrae}, {Bellas-Velidis},
  {Benson}, {Berthier}, {Blomme}, {Brugaletta}, {Burgess}, {Busso}, {Carry},
  {Cellino}, {Cheek}, {Clementini}, {Damerdji}, {Davidson}, {Delchambre},
  {Dell'Oro}, {Fern{\'a}ndez-Hern{\'a}ndez}, {Galluccio}, {Garc{\'\i}a-Lario},
  {Garcia-Reinaldos}, {Gonz{\'a}lez-N{\'u}{\~n}ez}, {Gosset}, {Haigron},
  {Halbwachs}, {Hambly}, {Harrison}, {Hatzidimitriou}, {Heiter},
  {Hern{\'a}ndez}, {Hestroffer}, {Hodgkin}, {Holl}, {Jan{\ss}en}, {Jevardat de
  Fombelle}, {Jordan}, {Krone-Martins}, {Lanzafame}, {L{\"o}ffler}, {Lorca},
  {Manteiga}, {Marchal}, {Marrese}, {Moitinho}, {Mora}, {Muinonen}, {Osborne},
  {Pancino}, {Pauwels}, {Petit}, {Recio-Blanco}, {Richards}, {Riello},
  {Rimoldini}, {Robin}, {Roegiers}, {Rybizki}, {Sarro}, {Siopis}, {Smith},
  {Sozzetti}, {Ulla}, {Utrilla}, {van Leeuwen}, {van Reeven}, {Abbas}, {Abreu
  Aramburu}, {Accart}, {Aerts}, {Aguado}, {Ajaj}, {Altavilla}, {{\'A}lvarez},
  {{\'A}lvarez Cid-Fuentes}, {Alves}, {Anderson}, {Anglada Varela}, {Antoja},
  {Audard}, {Baines}, {Baker}, {Balaguer-N{\'u}{\~n}ez}, {Balbinot}, {Balog},
  {Barache}, {Barbato}, {Barros}, {Barstow}, {Bartolom{\'e}}, {Bassilana},
  {Bauchet}, {Baudesson-Stella}, {Becciani}, {Bellazzini}, {Bernet}, {Bertone},
  {Bianchi}, {Blanco-Cuaresma}, {Boch}, {Bombrun}, {Bossini}, {Bouquillon},
  {Bragaglia}, {Bramante}, {Breedt}, {Bressan}, {Brouillet}, {Bucciarelli},
  {Burlacu}, {Busonero}, {Butkevich}, {Buzzi}, {Caffau}, {Cancelliere},
  {C{\'a}novas}, {Cantat-Gaudin}, {Carballo}, {Carlucci}, {Carnerero},
  {Carrasco}, {Casamiquela}, {Castellani}, {Castro-Ginard}, {Castro Sampol},
  {Chaoul}, {Charlot}, {Chemin}, {Chiavassa}, {Cioni}, {Comoretto}, {Cooper},
  {Cornez}, {Cowell}, {Crifo}, {Crosta}, {Crowley}, {Dafonte}, {Dapergolas},
  {David}, {David}, {de Laverny}, {De Luise}, {De March}, {De Ridder}, {de
  Souza}, {de Teodoro}, {de Torres}, {del Peloso}, {del Pozo}, {Delbo},
  {Delgado}, {Delgado}, {Delisle}, {Di Matteo}, {Diakite}, {Diener},
  {Distefano}, {Dolding}, {Eappachen}, {Edvardsson}, {Enke}, {Esquej}, {Fabre},
  {Fabrizio}, {Faigler}, {Fedorets}, {Fernique}, {Fienga}, {Figueras},
  {Fouron}, {Fragkoudi}, {Fraile}, {Franke}, {Gai}, {Garabato},
  {Garcia-Gutierrez}, {Garc{\'\i}a-Torres}, {Garofalo}, {Gavras}, {Gerlach},
  {Geyer}, {Giacobbe}, {Gilmore}, {Girona}, {Giuffrida}, {Gomel}, {Gomez},
  {Gonzalez-Santamaria}, {Gonz{\'a}lez-Vidal}, {Granvik},
  {Guti{\'e}rrez-S{\'a}nchez}, {Guy}, {Hauser}, {Haywood}, {Helmi}, {Hidalgo},
  {Hilger}, {H{\l}adczuk}, {Hobbs}, {Holland}, {Huckle}, {Jasniewicz},
  {Jonker}, {Juaristi Campillo}, {Julbe}, {Karbevska}, {Kervella}, {Khanna},
  {Kochoska}, {Kontizas}, {Kordopatis}, {Korn}, {Kostrzewa-Rutkowska},
  {Kruszy{\'n}ska}, {Lambert}, {Lanza}, {Lasne}, {Le Campion}, {Le Fustec},
  {Lebreton}, {Lebzelter}, {Leccia}, {Leclerc}, {Lecoeur-Taibi}, {Liao},
  {Licata}, {Lindstr{\o}m}, {Lister}, {Livanou}, {Lobel}, {Madrero Pardo},
  {Managau}, {Mann}, {Marchant}, {Marconi}, {Marcos Santos}, {Marinoni},
  {Marocco}, {Marshall}, {Martin Polo}, {Mart{\'\i}n-Fleitas}, {Masip},
  {Massari}, {Mastrobuono-Battisti}, {Mazeh}, {McMillan}, {Messina},
  {Michalik}, {Millar}, {Mints}, {Molina}, {Molinaro}, {Moln{\'a}r},
  {Montegriffo}, {Mor}, {Morbidelli}, {Morel}, {Morris}, {Mulone}, {Munoz},
  {Muraveva}, {Murphy}, {Musella}, {Noval}, {Ord{\'e}novic}, {Orr{\`u}},
  {Osinde}, {Pagani}, {Pagano}, {Palaversa}, {Palicio}, {Panahi}, {Pawlak},
  {Pe{\~n}alosa Esteller}, {Penttil{\"a}}, {Piersimoni}, {Pineau}, {Plachy},
  {Plum}, {Poggio}, {Poretti}, {Poujoulet}, {Pr{\v{s}}a}, {Pulone}, {Racero},
  {Ragaini}, {Rainer}, {Raiteri}, {Rambaux}, {Ramos}, {Ramos-Lerate}, {Re
  Fiorentin}, {Regibo}, {Reyl{\'e}}, {Ripepi}, {Riva}, {Rixon}, {Robichon},
  {Robin}, {Roelens}, {Rohrbasser}, {Romero-G{\'o}mez}, {Rowell}, {Royer},
  {Rybicki}, {Sadowski}, {Sagrist{\`a} Sell{\'e}s}, {Sahlmann}, {Salgado},
  {Salguero}, {Samaras}, {Sanchez Gimenez}, {Sanna}, {Santove{\~n}a},
  {Sarasso}, {Schultheis}, {Sciacca}, {Segol}, {Segovia}, {S{\'e}gransan},
  {Semeux}, {Shahaf}, {Siddiqui}, {Siebert}, {Siltala}, {Slezak}, {Smart},
  {Solano}, {Solitro}, {Souami}, {Souchay}, {Spagna}, {Spoto}, {Steele},
  {Steidelm{\"u}ller}, {Stephenson}, {S{\"u}veges}, {Szabados}, {Szegedi-Elek},
  {Taris}, {Tauran}, {Taylor}, {Teixeira}, {Thuillot}, {Tonello}, {Torra},
  {Torra}, {Turon}, {Unger}, {Vaillant}, {van Dillen}, {Vanel}, {Vecchiato},
  {Viala}, {Vicente}, {Voutsinas}, {Weiler}, {Wevers}, {Wyrzykowski}, {Yoldas},
  {Yvard}, {Zhao}, {Zorec}, {Zucker}, {Zurbach}, \& {Zwitter}}]{gaiaedr3}
---. 2021, \aap, 649, A1

\bibitem[{{Hasegawa} {et~al.}(2021){Hasegawa}, {Haworth}, {Hoadley}, {Kim},
  {Goto}, {Juzikenaite}, {Turner}, {Pascucci}, \& {Hamden}}]{hasegawa2021}
{Hasegawa}, Y., {Haworth}, T.~J., {Hoadley}, K., {Kim}, J.~S., {Goto}, H.,
  {Juzikenaite}, A., {Turner}, N.~J., {Pascucci}, I., \& {Hamden}, E.~T. 2021,
  arXiv e-prints, arXiv:2112.02831

\bibitem[{{Hashimoto} {et~al.}(2021){Hashimoto}, {Muto}, {Dong}, {Liu}, {van
  der Marel}, {Francis}, {Hasegawa}, \& {Tsukagoshi}}]{hashimoto2021}
{Hashimoto}, J., {Muto}, T., {Dong}, R., {Liu}, H.~B., {van der Marel}, N.,
  {Francis}, L., {Hasegawa}, Y., \& {Tsukagoshi}, T. 2021, \apj, 911, 5

\bibitem[{{Ishihara} {et~al.}(2010){Ishihara}, {Onaka}, {Kataza}, {Salama},
  {Alfageme}, {Cassatella}, {Cox}, {Garc{\'\i}a-Lario}, {Stephenson}, {Cohen},
  {Fujishiro}, {Fujiwara}, {Hasegawa}, {Ita}, {Kim}, {Matsuhara}, {Murakami},
  {M{\"u}ller}, {Nakagawa}, {Ohyama}, {Oyabu}, {Pyo}, {Sakon}, {Shibai},
  {Takita}, {Tanab{\'e}}, {Uemizu}, {Ueno}, {Usui}, {Wada}, {Watarai},
  {Yamamura}, \& {Yamauchi}}]{AKARI2010}
{Ishihara}, D., {et~al.} 2010, \aap, 514, A1

\bibitem[{{Kanagawa} {et~al.}(2021){Kanagawa}, {Hashimoto}, {Muto},
  {Tsukagoshi}, {Takahashi}, {Hasegawa}, {Konishi}, {Nomura}, {Liu}, {Dong},
  {Kataoka}, {Momose}, {Ono}, {Sitko}, {Takami}, \& {Tomida}}]{kanagawa2021}
{Kanagawa}, K.~D., {et~al.} 2021, \apj, 909, 212

\bibitem[{{Kudo} {et~al.}(2018){Kudo}, {Hashimoto}, {Muto}, {Liu}, {Dong},
  {Hasegawa}, {Tsukagoshi}, \& {Konishi}}]{Kudo2018}
{Kudo}, T., {Hashimoto}, J., {Muto}, T., {Liu}, H.~B., {Dong}, R., {Hasegawa},
  Y., {Tsukagoshi}, T., \& {Konishi}, M. 2018, \apjl, 868, L5

\bibitem[{{Liu}(2019)}]{baobab2019}
{Liu}, H.~B. 2019, \apjl, 877, L22

\bibitem[{{Manara} {et~al.}(2017){Manara}, {Testi}, {Herczeg}, {Pascucci},
  {Alcal{\'a}}, {Natta}, {Antoniucci}, {Fedele}, {Mulders}, {Henning},
  {Mohanty}, {Prusti}, \& {Rigliaco}}]{manara2017}
{Manara}, C.~F., {et~al.} 2017, \aap, 604, A127

\bibitem[{{Mangum} \& {Shirley}(2015)}]{mangum2015}
{Mangum}, J.~G., \& {Shirley}, Y.~L. 2015, \pasp, 127, 266

\bibitem[{{Marino} {et~al.}(2015){Marino}, {Perez}, \& {Casassus}}]{marino2015}
{Marino}, S., {Perez}, S., \& {Casassus}, S. 2015, \apjl, 798, L44

\bibitem[{{Mayama} {et~al.}(2018){Mayama}, {Akiyama}, {Pani{\'c}}, {Miley},
  {Tsukagoshi}, {Muto}, {Dong}, {de Leon}, {Mizuki}, {Oh}, {Hashimoto}, {Sai},
  {Currie}, {Takami}, {Grady}, {Hayashi}, {Tamura}, \& {Inutsuka}}]{mayama2018}
{Mayama}, S., {et~al.} 2018, \apjl, 868, L3

\bibitem[{{McMullin} {et~al.}(2007){McMullin}, {Waters}, {Schiebel}, {Young},
  \& {Golap}}]{macmullin2007}
{McMullin}, J.~P., {Waters}, B., {Schiebel}, D., {Young}, W., \& {Golap}, K.
  2007, in Astronomical Society of the Pacific Conference Series, Vol. 376,
  Astronomical Data Analysis Software and Systems XVI, ed. R.~A. {Shaw},
  F.~{Hill}, \& D.~J. {Bell}, 127

\bibitem[{{Neugebauer} {et~al.}(1984){Neugebauer}, {Habing}, {van Duinen},
  {Aumann}, {Baud}, {Beichman}, {Beintema}, {Boggess}, {Clegg}, {de Jong},
  {Emerson}, {Gautier}, {Gillett}, {Harris}, {Hauser}, {Houck}, {Jennings},
  {Low}, {Marsden}, {Miley}, {Olnon}, {Pottasch}, {Raimond}, {Rowan-Robinson},
  {Soifer}, {Walker}, {Wesselius}, \& {Young}}]{IRAS1984}
{Neugebauer}, G., {et~al.} 1984, \apjl, 278, L1

\bibitem[{{Pinilla} {et~al.}(2012){Pinilla}, {Benisty}, \&
  {Birnstiel}}]{pinilla2012}
{Pinilla}, P., {Benisty}, M., \& {Birnstiel}, T. 2012, \aap, 545, A81

\bibitem[{{Pinilla} {et~al.}(2018){Pinilla}, {Tazzari}, {Pascucci}, {Youdin},
  {Garufi}, {Manara}, {Testi}, {van der Plas}, {Barenfeld}, {Canovas}, {Cox},
  {Hendler}, {P{\'e}rez}, \& {van der Marel}}]{pinilla2018}
{Pinilla}, P., {et~al.} 2018, \apj, 859, 32

\bibitem[{{Pontoppidan} {et~al.}(2008){Pontoppidan}, {Blake}, {van Dishoeck},
  {Smette}, {Ireland}, \& {Brown}}]{pontoppidan2008}
{Pontoppidan}, K.~M., {Blake}, G.~A., {van Dishoeck}, E.~F., {Smette}, A.,
  {Ireland}, M.~J., \& {Brown}, J. 2008, \apj, 684, 1323

\bibitem[{{Qi} {et~al.}(2011){Qi}, {D'Alessio}, {{\"O}berg}, {Wilner},
  {Hughes}, {Andrews}, \& {Ayala}}]{Qi2011}
{Qi}, C., {D'Alessio}, P., {{\"O}berg}, K.~I., {Wilner}, D.~J., {Hughes},
  A.~M., {Andrews}, S.~M., \& {Ayala}, S. 2011, \apj, 740, 84

\bibitem[{{Rosenfeld} {et~al.}(2014){Rosenfeld}, {Chiang}, \&
  {Andrews}}]{rosenfeld2014}
{Rosenfeld}, K.~A., {Chiang}, E., \& {Andrews}, S.~M. 2014, \apj, 782, 62

\bibitem[{{Sheikhnezami} {et~al.}(2012){Sheikhnezami}, {Fendt}, {Porth},
  {Vaidya}, \& {Ghanbari}}]{Sheikhnezami2012}
{Sheikhnezami}, S., {Fendt}, C., {Porth}, O., {Vaidya}, B., \& {Ghanbari}, J.
  2012, \apj, 757, 65

\bibitem[{{Siess} {et~al.}(2000){Siess}, {Dufour}, \& {Forestini}}]{siess2000}
{Siess}, L., {Dufour}, E., \& {Forestini}, M. 2000, \aap, 358, 593

\bibitem[{{Siess} {et~al.}(1997){Siess}, {Forestini}, \&
  {Dougados}}]{siess1997}
{Siess}, L., {Forestini}, M., \& {Dougados}, C. 1997, \aap, 324, 556

\bibitem[{{Soon} {et~al.}(2019){Soon}, {Momose}, {Muto}, {Tsukagoshi},
  {Kataoka}, {Hanawa}, {Fukagawa}, {Saigo}, \& {Shibai}}]{soon2019}
{Soon}, K.-L., {Momose}, M., {Muto}, T., {Tsukagoshi}, T., {Kataoka}, A.,
  {Hanawa}, T., {Fukagawa}, M., {Saigo}, K., \& {Shibai}, H. 2019, \pasj, 71,
  124

\bibitem[{{Suzuki} {et~al.}(2016){Suzuki}, {Ogihara}, {Morbidelli}, {Crida}, \&
  {Guillot}}]{suzuki2016}
{Suzuki}, T.~K., {Ogihara}, M., {Morbidelli}, A., {Crida}, A., \& {Guillot}, T.
  2016, \aap, 596, A74

\bibitem[{{Tazaki} {et~al.}(2016){Tazaki}, {Tanaka}, {Okuzumi}, {Kataoka}, \&
  {Nomura}}]{tazaki2016}
{Tazaki}, R., {Tanaka}, H., {Okuzumi}, S., {Kataoka}, A., \& {Nomura}, H. 2016,
  \apj, 823, 70

\bibitem[{{Teague}(2020)}]{teague2020}
{Teague}, R. 2020, {richteague/keplerian\_mask: Initial Release},
  doi:10.5281/zenodo.4321137

\bibitem[{{Teague} \& {Foreman-Mackey}(2018)}]{better2018}
{Teague}, R., \& {Foreman-Mackey}, D. 2018, {Bettermoments: A Robust Method To
  Measure Line Centroids}, doi:10.5281/zenodo.1419754

\bibitem[{{Tsukagoshi} {et~al.}(2019){Tsukagoshi}, {Momose}, {Kitamura},
  {Saito}, {Kawabe}, {Andrews}, {Wilner}, {Kudo}, {Hashimoto}, {Ohashi}, \&
  {Tamura}}]{tsukagoshi2019}
{Tsukagoshi}, T., {et~al.} 2019, \apj, 871, 5

\bibitem[{{Tsukamoto} {et~al.}(2015{\natexlab{a}}){Tsukamoto}, {Iwasaki},
  {Okuzumi}, {Machida}, \& {Inutsuka}}]{tsukamoto2015b}
{Tsukamoto}, Y., {Iwasaki}, K., {Okuzumi}, S., {Machida}, M.~N., \& {Inutsuka},
  S. 2015{\natexlab{a}}, \apjl, 810, L26

\bibitem[{{Tsukamoto} {et~al.}(2015{\natexlab{b}}){Tsukamoto}, {Iwasaki},
  {Okuzumi}, {Machida}, \& {Inutsuka}}]{tsukamoto2015a}
---. 2015{\natexlab{b}}, \mnras, 452, 278

\bibitem[{{Ueda} {et~al.}(2020){Ueda}, {Kataoka}, \& {Tsukagoshi}}]{ueda2020}
{Ueda}, T., {Kataoka}, A., \& {Tsukagoshi}, T. 2020, \apj, 893, 125

\bibitem[{{van der Marel} {et~al.}(2013){van der Marel}, {van Dishoeck},
  {Bruderer}, {Birnstiel}, {Pinilla}, {Dullemond}, {van Kempen}, {Schmalzl},
  {Brown}, {Herczeg}, {Mathews}, \& {Geers}}]{ninke2013}
{van der Marel}, N., {et~al.} 2013, Science, 340, 1199

\bibitem[{{van der Marel} {et~al.}(2018){van der Marel}, {Williams}, {Ansdell},
  {Manara}, {Miotello}, {Tazzari}, {Testi}, {Hogerheijde}, {Bruderer}, {van
  Terwisga}, \& {van Dishoeck}}]{ninke2018}
---. 2018, \apj, 854, 177

\bibitem[{{Winston} {et~al.}(2012){Winston}, {Cox}, {Prusti}, {Mer{\'\i}n},
  {Ribas}, {Royer}, {Vavrek}, {Puga}, {Andr{\'e}}, {Men'shchikov},
  {K{\"o}nyves}, {K{\'o}sp{\'a}l}, {Alves de Oliveira}, {Pilbratt}, \&
  {Waelkens}}]{Herschel2012}
{Winston}, E., {et~al.} 2012, \aap, 545, A145

\bibitem[{{Wright} {et~al.}(2010){Wright}, {Eisenhardt}, {Mainzer}, {Ressler},
  {Cutri}, {Jarrett}, {Kirkpatrick}, {Padgett}, {McMillan}, {Skrutskie},
  {Stanford}, {Cohen}, {Walker}, {Mather}, {Leisawitz}, {Gautier}, {McLean},
  {Benford}, {Lonsdale}, {Blain}, {Mendez}, {Irace}, {Duval}, {Liu}, {Royer},
  {Heinrichsen}, {Howard}, {Shannon}, {Kendall}, {Walsh}, {Larsen}, {Cardon},
  {Schick}, {Schwalm}, {Abid}, {Fabinsky}, {Naes}, \& {Tsai}}]{WISE2010}
{Wright}, E.~L., {et~al.} 2010, \aj, 140, 1868

\bibitem[{{Zhang} {et~al.}(2020){Zhang}, {Schwarz}, \& {Bergin}}]{Zhang2020}
{Zhang}, K., {Schwarz}, K.~R., \& {Bergin}, E.~A. 2020, \apjl, 891, L17

\end{thebibliography}
\bibliographystyle{apj}

\end{document}